\documentclass[useAMS,usenatbib]{mn2e}
\usepackage{epsfig}

\usepackage{lipsum}
\usepackage{lscape}
\usepackage{longtable,amsmath}
\usepackage{booktabs}
\usepackage{marvosym}
\usepackage{pdflscape}
\usepackage[none]{hyphenat}
\usepackage[T1]{fontenc}
\usepackage{aecompl}

\title[Star Cluster Scale Height]{Properties of Star Clusters - II: Scale Height Evolution of Clusters}

\author[Buckner \& Froebrich]{Anne~S.M. Buckner$^{1}$\thanks{E-mail:
asmb2@kent.ac.uk}, Dirk Froebrich$^{1}$\thanks{E-mail: df@star.kent.ac.uk}\\
$^{1}$ Centre for Astrophysics and Planetary Science, University of Kent,
Canterbury, CT2 7NH, United Kingdom }

\begin{document}

\date{Accepted. Received.}

\pagerange{\pageref{firstpage}--\pageref{lastpage}} \pubyear{2013}

\maketitle

\label{firstpage}

\begin{abstract}

Until now it has been impossible to observationally measure how star cluster
scale height evolves beyond 1\,Gyr as only small samples have been available.
Here we establish a novel method to determine the scale height of a cluster
sample using modelled distributions and Kolmogorov-Smirnov tests. This allows us
to determine the scale height with a 25\,\% accuracy for samples of 38 clusters
or more. We apply our method to investigate the temporal evolution of cluster
scale height, using homogeneously selected sub-samples of Kharchenko et al.
(MWSC), Dias et al. (DAML02), WEBDA, and Froebrich et al. (FSR).

We identify a linear relationship between scale height and $log(age/yr)$ of 
clusters, considerably different from field stars. The scale height increases
from about 40\,pc at 1\,Myr to 75\,pc at 1\,Gyr, most likely due to internal 
evolution and external scattering events. After 1\,Gyr, there is a marked 
change of the behaviour, with the scale height linearly increasing with 
$log(age/yr)$ to about 550\,pc at 3.5\,Gyr. The most likely interpretation is 
that the surviving clusters are only observable because they have been 
scattered away from the mid-plane in their past. A detailed understanding of
this observational evidence can only be achieved with numerical simulations of 
the evolution of cluster samples in the Galactic Disk.

Furthermore, we find a weak trend of an age-independent increase in scale 
height with galactocentric distance. There are no significant temporal or 
spatial variations of the cluster distribution zero point. We determine the
Sun's vertical displacement from the Galactic Plane as
$Z_{\odot}=18.5\pm1.2$\,pc.

\end{abstract}


\section{Introduction}\label{intro}

Star clusters act as tracers of stellar and Galactic evolution and are the
building blocks of the Galaxy. The majority of stars in the Galaxy are formed in
open clusters \citep{Lada2003}, and as such it is important to determine
fundamental properties of both individual clusters (e.g. age, distance,
reddening, metallicity), and large cluster samples (e.g. spatial distribution
within and across the Galactic Plane, i.e. the scale height).

Open clusters are formed in Giant Molecular Clouds (GMCs) and can remain
embedded for up to 10\,Myrs. As an embedded cluster evolves, stellar feedback
(i.e. stellar winds, jets, outflows, supernovae) influences the gas internal to
the cluster. The resulting (radiative) pressure drives the gas outwards, it
eventually disperses and a bound open cluster might emerge. During this phase
the mass loss (from gas) will cause the majority of embedded clusters to be
disrupted and dissolve into the field, with only about 5\,\% emerging and
evolving to become bound open clusters (e.g. \citet{Lada2003}). Once emerged,
clusters face dissolution into the field via dynamical mass segregation, tidal
stripping and disruption from gravitational interactions with e.g. GMCs.
Estimated disruption time-scales are 10\,--\,40\,Myr, correlating with the
cluster's distance from the Galactic Centre (e.g. \citet{2006MNRAS.373..752G}).
Few clusters survive to 1\,Gyr which is highlighted by the lack of older
clusters observed in the solar neighbourhood in comparison to younger clusters.

To fully understand open cluster behaviour on a Galactic scale, it is important
to begin to build up an observational picture of the evolution of scale height
with cluster age. Previous works have shown that older clusters (age above
1\,Gyr) have a typical scale height of 375\,pc \citep{2010MNRAS.409.1281F},
significantly larger than their younger counterparts. Unfortunately, methods to
determine the scale height are only applicable to larger sample sizes and fail
in the case of small samples of rare old clusters. Thus, it has been difficult,
observationally, to investigate the evolution of cluster scale height in smaller
age bins, especially for the rare old objects.

Additional difficulties lie in the nature of open cluster catalogues (e.g.
WEBDA, or DAML02 \citep{2002A&A...389..871D}) as fundamental cluster parameters
are often compiled from the literature and are hence not homogeneously
determined. For example \citet{2010MNRAS.409.1281F} found that FSR\,1716 has a
distance of 7.0\,kpc and an age of 2\,Gyr, whereas \citet{2008A&A...491..767B}
determined the cluster to have a distance/age of either 0.8\,kpc$/$7\,Gyr or
2.3\,kpc$/$12\,Gyr, respectively. Note that the differences in this case mainly
arise from using different metallicities when estimating the parameters and
interpreting features along the isochrone differently, or the whole cluster as a
globular or open cluster. However, it serves as an example that homogeneously
derived cluster lists, where any uncertainties in the determined values are
systematic, are essential for a comprehensive analysis of large cluster samples.

In this series of papers we aim to homogeneously and statistically investigate
the fundamental properties and large scale distribution of open clusters in the
Galaxy. In \citet{2013MNRAS.436.1465B} (Paper\,I, hereafter) we established a
foreground star counting technique as a distance measurement and presented an
automatic calibration and optimisation method for use on large samples of
clusters with Near-Infrared (NIR) photometric data only. We combined this method
with colour excess calculations to determine distances and extinctions of
objects in the FSR cluster sample from \citet{2007MNRAS.374..399F} and
investigated the H-band extinction per kpc distance in the Galactic Disk as a
function of Galactic longitude. In total, we determined distance estimates to
771, and extinctions values for 775, open cluster candidates from the FSR list.

In this paper we investigate the relationship between scale height and cluster
age. We will use our novel approach to calculate cluster scale heights, which
can be applied to small sample sizes. We begin by building upon the work of
\citet{2010MNRAS.409.1281F}, who determined the ages of the 'old' ($>$100\,Myr)
FSR cluster candidates, by homogeneously fitting isochrones to derive the ages
of our FSR sub-sample and further refine their determined distance and
extinction values. We follow this with a comprehensive analysis of the scale
height of clusters in the homogeneous MWSC catalogue by
\citet{2013A&A...558A..53K}, the DAML02 list by \citet{2002A&A...389..871D} and
the WEBDA database.

This paper is structured as follows. In Sec.\,\ref{sect_age} we present our
cluster sample and subsequent age analysis. Section\,\ref{sect_sh} introduces
our novel scale height approach. The results of our scale height and age
analyses are discussed in Sec.\,\ref{sect_results}. Our conclusions are
presented in Sec.\,\ref{sect_conclusion}.

\section{Cluster Samples}\label{sect_samp}

\begin{figure*}
\includegraphics[width=8.6cm,angle=0]{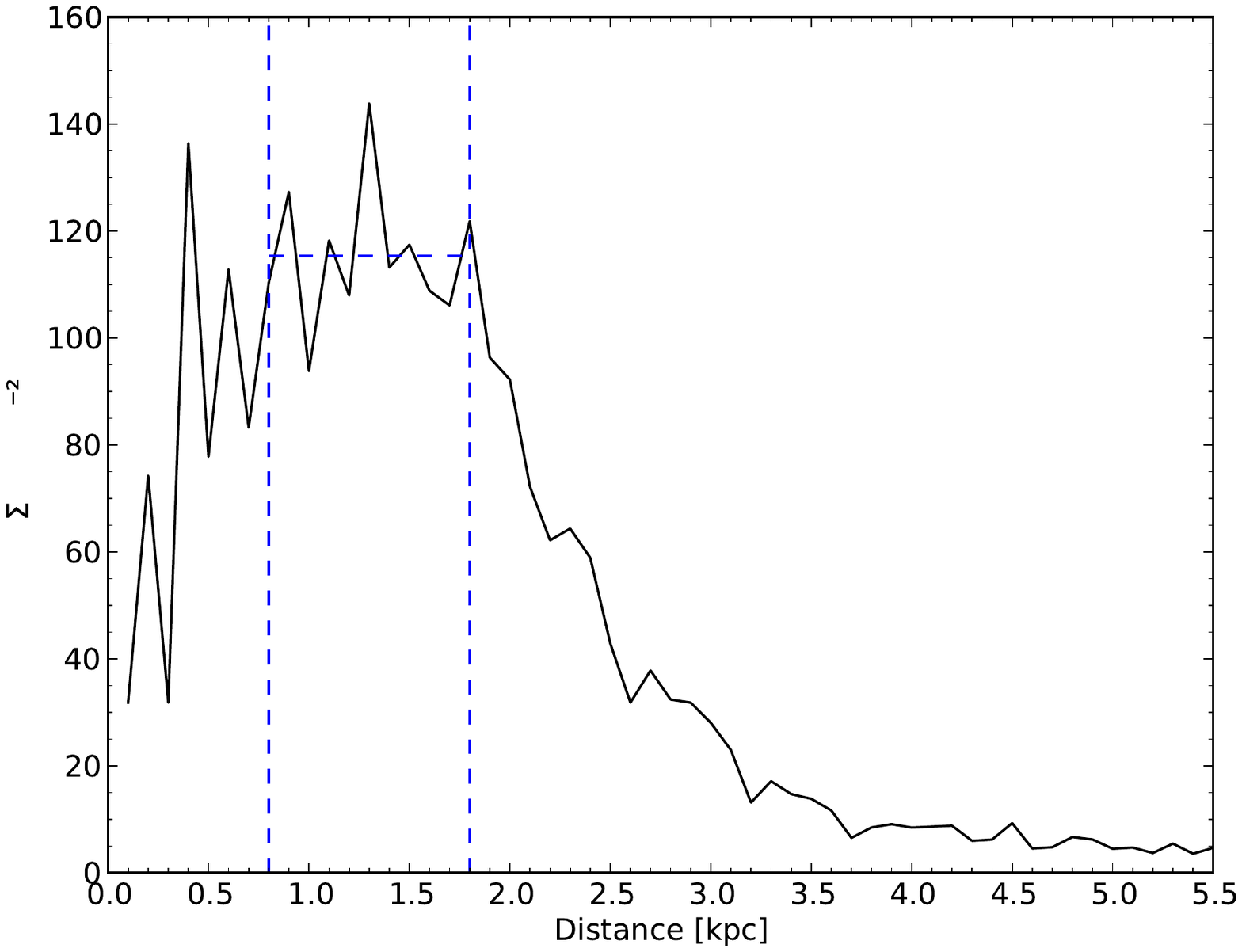} \hfill
\includegraphics[width=8.6cm,angle=0]{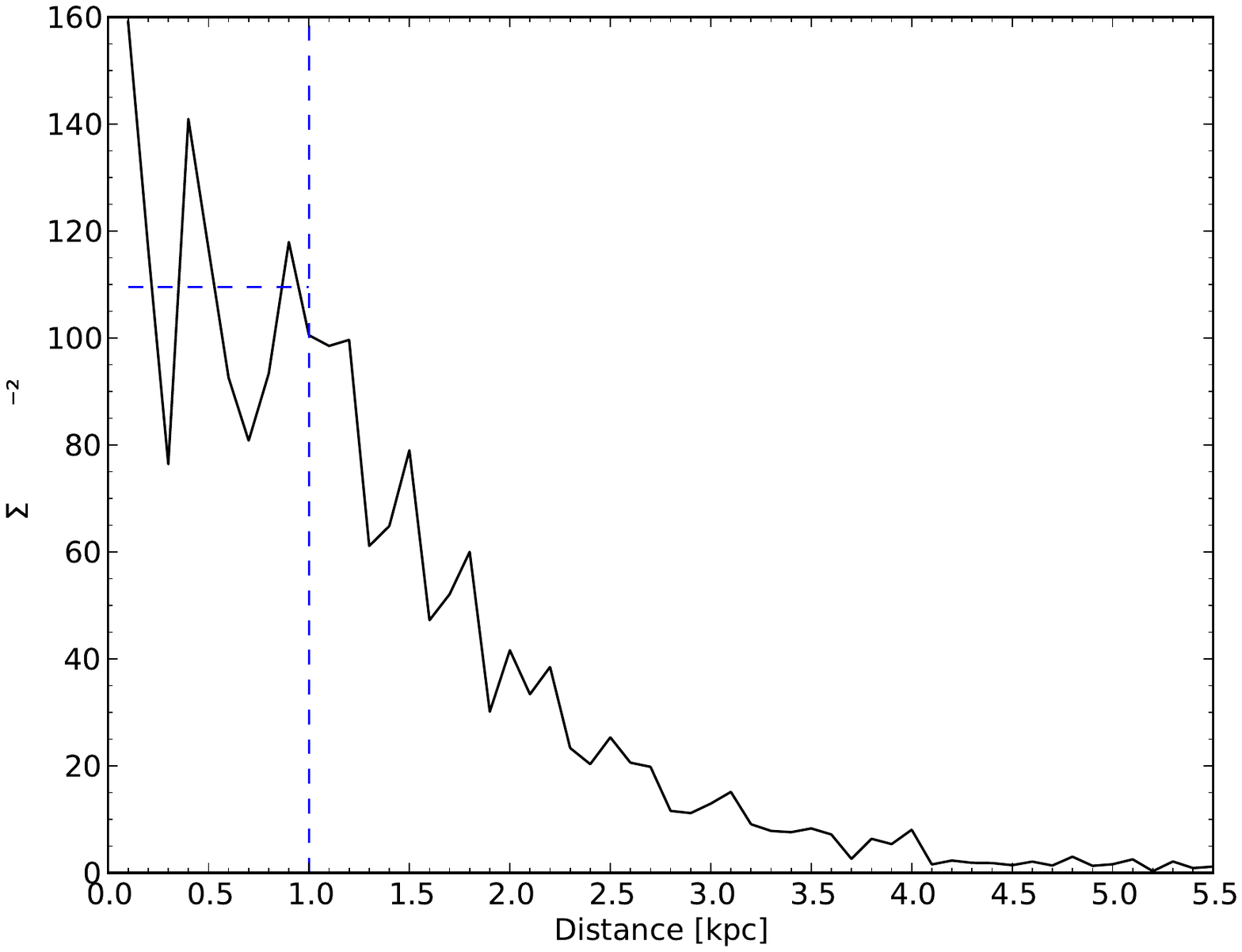} \\
\includegraphics[width=8.6cm,angle=0]{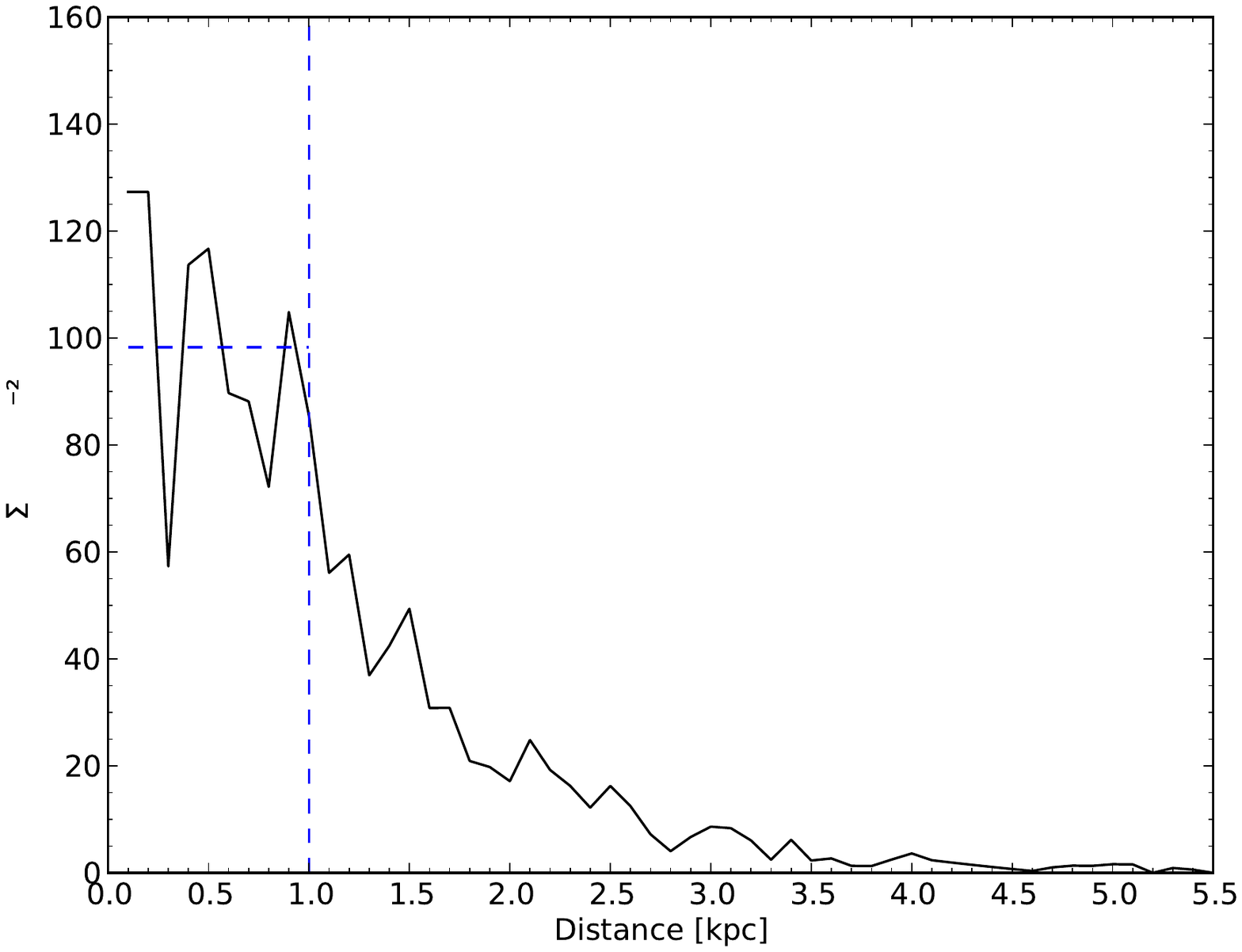} \hfill
\includegraphics[width=8.6cm,angle=0]{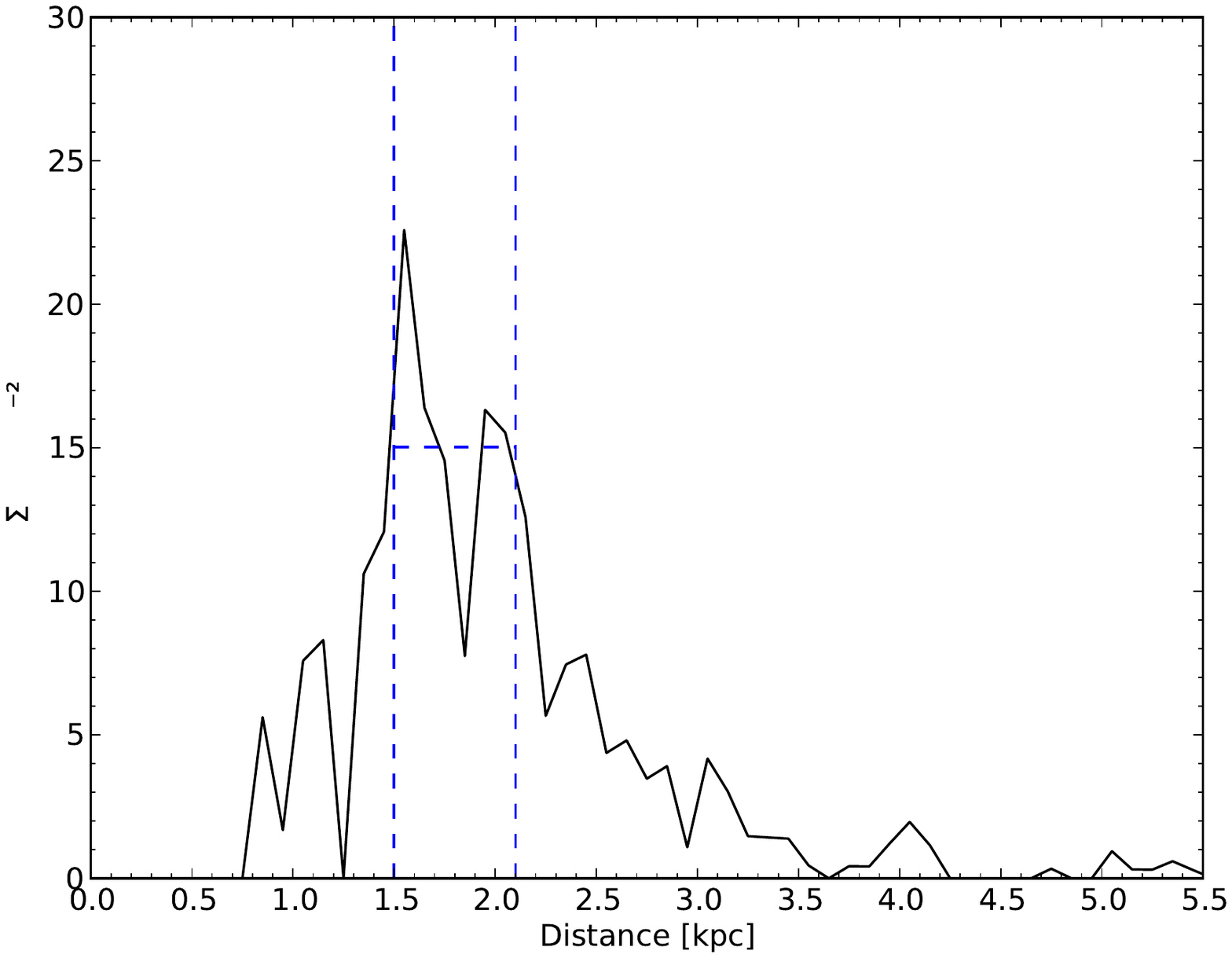}

\caption{\label{samples}  Surface density distribution of clusters as a function
of distance in the samples investigated in our work (top-left: CS\,1 --
Kharchenko; top-right: CS\,2 -- Dias; bottom-left: CS\,3 -- WEBDA; bottom-right:
CS\,4 -- FSR). In each panel the vertical dashed line(s) indicate the region
where we consider the sample to have a homogeneous completeness and the
horizontal dashed line indicates the surface density in this region. }.

\end{figure*}

In the latter part of this paper we aim to investigate the temporal and spatial
scale height evolution of samples of clusters in detail. Ideally we require a
variety of samples/catalogues to identify potential selection effects in them.
Most importantly, however, we require a large number of clusters with a
significant age spread and an extended distribution in the Galactic Plane to
investigate positional variations of the scale height. There are four obvious
choices of cluster samples (CS), each with its own advantages and disadvantages:

\begin{enumerate}

\item {\bf CS\,1:} The MWSC catalogue by \citet{2013A&A...558A..53K}. This
catalogue was initially compiled from the literature (including many of the
clusters in our other CSs) and contains 3006 real clusters with an additional
few hundred that are flagged as either not real or duplicate entries. Using
their data-processing pipeline, the authors homogeneously re$-/$determined
distance, reddening, radii and age values for each object with isochrone fits
and data from the PPMXL and 2MASS catalogues (see source paper for further
details). Thus, any uncertainties in the cluster parameters are therefore
systematic and not caused by inconsistencies in the sample. To date this is the
most comprehensive, homogeneously derived star cluster catalogue available in
the literature, which coupled with its extensive spread of cluster ages, makes
it an invaluable resource.

From the catalogue we select only the real objects and exclude all the globular
clusters, associations and moving groups, as we are only interested in real
bound open clusters. For the purpose of our analysis moving groups, although
part of open cluster evolution, are considered no longer sufficiently bound to
be included. Objects flagged as 'Remnants' or 'Nebulous' are retained as they
are typically associated with very old and very young open clusters,
respectively.

We determine the completeness limit of the selected clusters by plotting the
distribution of the surface density of clusters ($\sum_{XY}$) against the
distance ($d_{XY}$) of the clusters from the Sun projected onto the Galactic
plane. Note the authors of the MWSC catalogue find a deficit of old open
clusters ($log(age/yr) > 9.2$) in the catalogue within 1\,kpc of Sun. The exact
cause for this is unknown, but it is reasonable to assume that this is due to
the natural evolution of clusters into a less-bound state, thus becoming too
large on the sky at short distances to be detectable. Taking the old cluster
deficit into account, CS\,1 is complete (or has at least a constant
completeness) at a distance range of 0.8\,--\,1.8\,kpc from the Sun for $ |b|
\le 90^{\circ}$, with an average surface density of 115\,clusters\,/\,kpc$^{2}$
(see top left panel of Fig.\,\ref{samples}). Thus we only select MWSC clusters
in this distance range to avoid any bias in the scale height determination later
on. This final selection leaves 960 clusters in the CS\,1 sample.

A recent study (Schmeja et al. 2014, MNRAS, accepted) based on 2MASS photometry
has identified a further 139, preferentially old, open clusters in the solar
neighbourhood at $|b| > 20^\circ$. Including them into the CS\,1 sample with the
same selections applied to MWSC, would increase the CS\,1 sample size by 79
objects. We refrain from doing this, since these new objects are exclusively at
large distances from the Galactic Plane, and would hence introduce a bias into
the sample.

\item {\bf CS\,2:} The current version of the
DAML02\footnote{http://www.astro.iag.usp.br/ocdb/} database by
\citet{2002A&A...389..871D}. This online database is compiled from the
literature and is regularly updated as new data becomes available. It contained
2174 objects at the time of writing. It is the largest open cluster database,
with the exception of the MWSC catalogue (of which it formed the basis of).
However, unlike the MWSC catalogue, the cluster parameters (distance, reddening,
age, etc.) have not been redetermined and remain as derived by the respective
authors of the literature. As such, the DAML02 database is inhomogeneous in
nature. However the extent of this inhomogeneity is unknown as the authors of
the parameters have analysed clusters on an individual basis i.e. extensively
and not as a collective where misinterpretation of data can be made due to the
systematic nature of the methods used to derive the parameters. For example, the
cluster Stephenson\,2 is a young massive cluster ($4 \times 10^{4} M_{\odot}$)
with 26 red supergiants at a distance of $5.8^{+1.9}_{-0.8} kpc$ and an age of
12\,--\,17\,Myr \citep{2007ApJ...671..781D}, but is listed as having a distance
of 1.1\,kpc with an age of 1\,Myr in MWSC . If the status of Stephenson\,2 as
being a young massive cluster is unknown (as was the case with MWSC in their
blind-data-processing pipeline), its colour-magnitude diagram can be 
misinterpreted. Thus, for a comprehensive scale height evolution analysis it is
of benefit to consider both cluster catalogues in order to compare the results
and to evaluate if there are systematic differences.

We select all clusters from the DAML02 database which have distance, reddening
and age value. Duplicate entries are identified as entries which had a
counterpart within 3.5$'$ and removed accordingly. The selected clusters are
determined to be complete, or have a homogeneous completeness, for up to a
1\,kpc radius from the Sun for $ |b| \le 90^{\circ}$ (see top-right panel in
Fig.\,\ref{samples}). The surface density of 110\,clusters\,/\,kpc$^{2}$ within
the 1\,kpc radius is comparable to the MWSC catalogue, i.e. CS\,1. The
selections leave 389 open clusters in the CS\,2 sample.

\item {\bf CS\,3:} The WEBDA\footnote{\tt http://www.univie.ac.at/webda/}
database based on \citet{1995ASSL..203..127M}. This online interactive database
of open clusters contains 1755 objects to date. WEBDA is compiled from the
literature, however it generally only includes high accuracy measurements,
compared to the more complete DAML02 database, thus making it a prudent choice
to include in our analysis in addition to both the DAML02 and MWSC data.

As for the first two cluster samples, we make a selection of objects which have
distance, reddening and age values. The selected clusters are determined to be
complete, have a homogeneous completeness, up to a 1\,kpc radius from the Sun
for $ |b| \le 90^{\circ}$ (see bottom left panel in Fig.\,\ref{samples}), with a
surface density of  98\,clusters\,/\,kpc$^{2}$. This is slightly less than the
values for CS\,1 and 2, but still comparable. The selections leaves 358 open
clusters in the CS\,3 sample.

\item {\bf CS\,4:} The FSR List by \citet{2007MNRAS.374..399F}. The authors of
this catalogue used 2MASS star density maps of the Milky Way across all Galactic
longitudes and within a Galactic latitude range of $|b| \le 20^{\circ}$ to
identify 1788 objects, including 87 globular clusters and 1021 previously
unknown open cluster candidates.

In Paper\,I we presented and calibrated automated methods to determine the
distances and extinctions to these star clusters using NIR photometry only and
foreground star counts. Uncertainties of better than 40\,\% where achieved for
the cluster distances , using a calibration sample with an intrinsic scatter of
30\,\%. We applied the method to the entire FSR list to determine distances and
extinctions for a sub-sample of 775 open cluster candidates with enough members,
of which 397 were new cluster candidates. Globular clusters were excluded as
they are prone to additional intrinsic effects that affect photometric quality
(e.g. central overcrowding) which could not be compensated for in our
calibration procedure (for full details see \citet{2013MNRAS.436.1465B}).

We aim to determine the ages of this FSR sub-sample using our data-processing
pipeline (see Sect. \ref{sect_isopip}). Clusters for which we were able to
accurately determine all 3 parameter values (age, distance, reddening), are then
selected and determined to have a homogeneous completeness at distances between
1.5\,--\,2.1\,kpc from the Sun for $|b| \le 20^{\circ}$, with a surface density
of 15\,clusters\,/\,kpc$^{2}$ (see bottom right panel of Fig.\,\ref{samples}).
The selections leave only 95 open clusters in the CS\,4 sample.

The above determined cluster surface density shows that this FSR catalogue
sub-sample  is only complete at the $\sim$\,13\,\% level. However, it extends
the cluster sample towards slightly larger distances, and contains a larger
fraction of older clusters, compared to the other samples. This is evident in
Fig.\,\ref{agedistributions} where we present the age distributions of all four
cluster samples. There the CSs\,1, 2, 3 show the normal trend that is expected
for samples selected as having a homogeneous completeness limit, i.e. a steeply
decreasing number of clusters with age. For CS\,4, however, the histogram is
more or less flat between 0.5 and 2.0\,Gyr. Furthermore,
Fig.\,\ref{agedistributions} also shows that the MWSC sample is the only sample
large enough to contain a sizable number of clusters older than 1\,--\,2\,Gyr,
or a large enough sample to potentially measure the age dependence of the scale
height of these objects.

\end{enumerate}

\begin{figure*}
\includegraphics[width=8.6cm,angle=0]{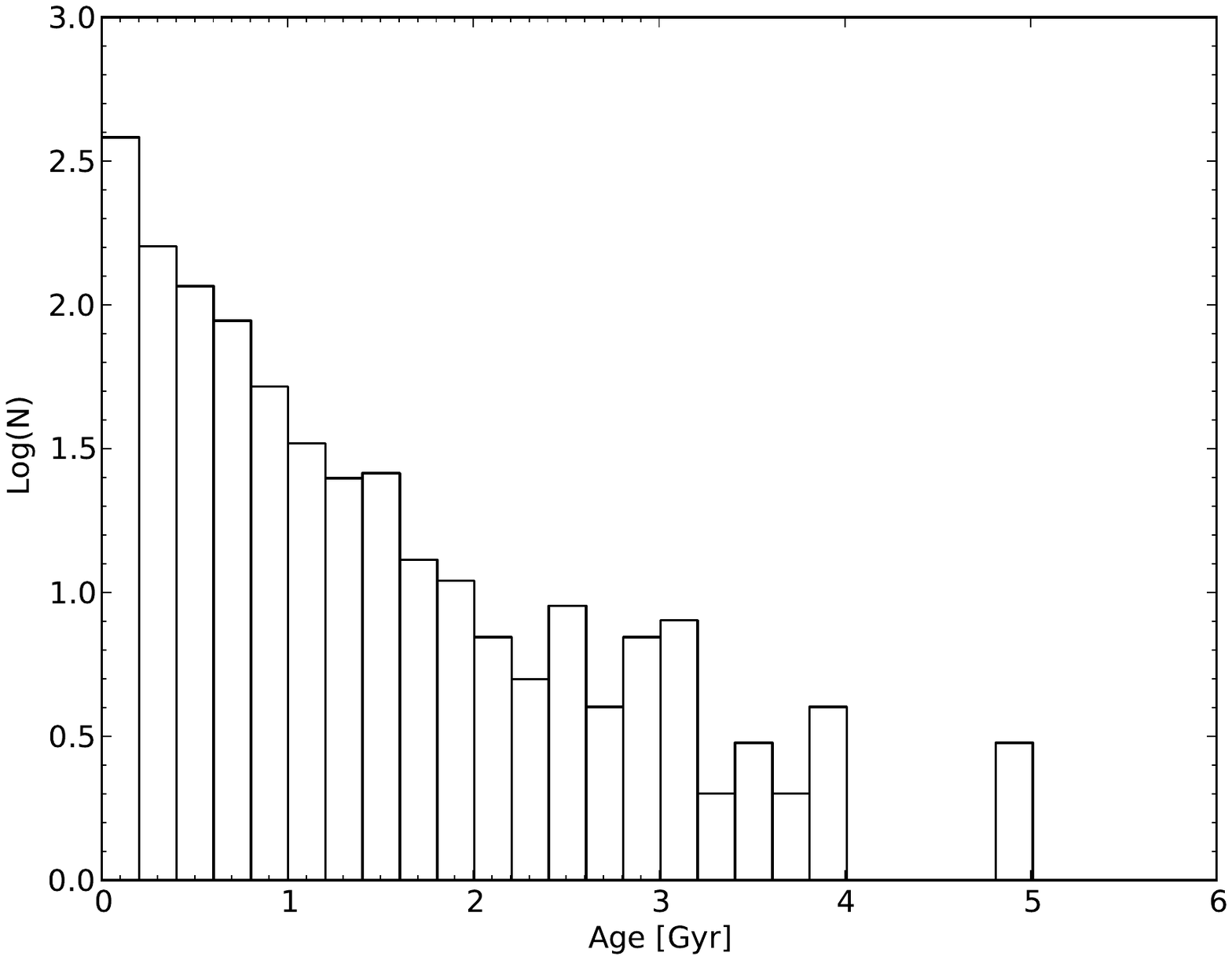} \hfill
\includegraphics[width=8.6cm,angle=0]{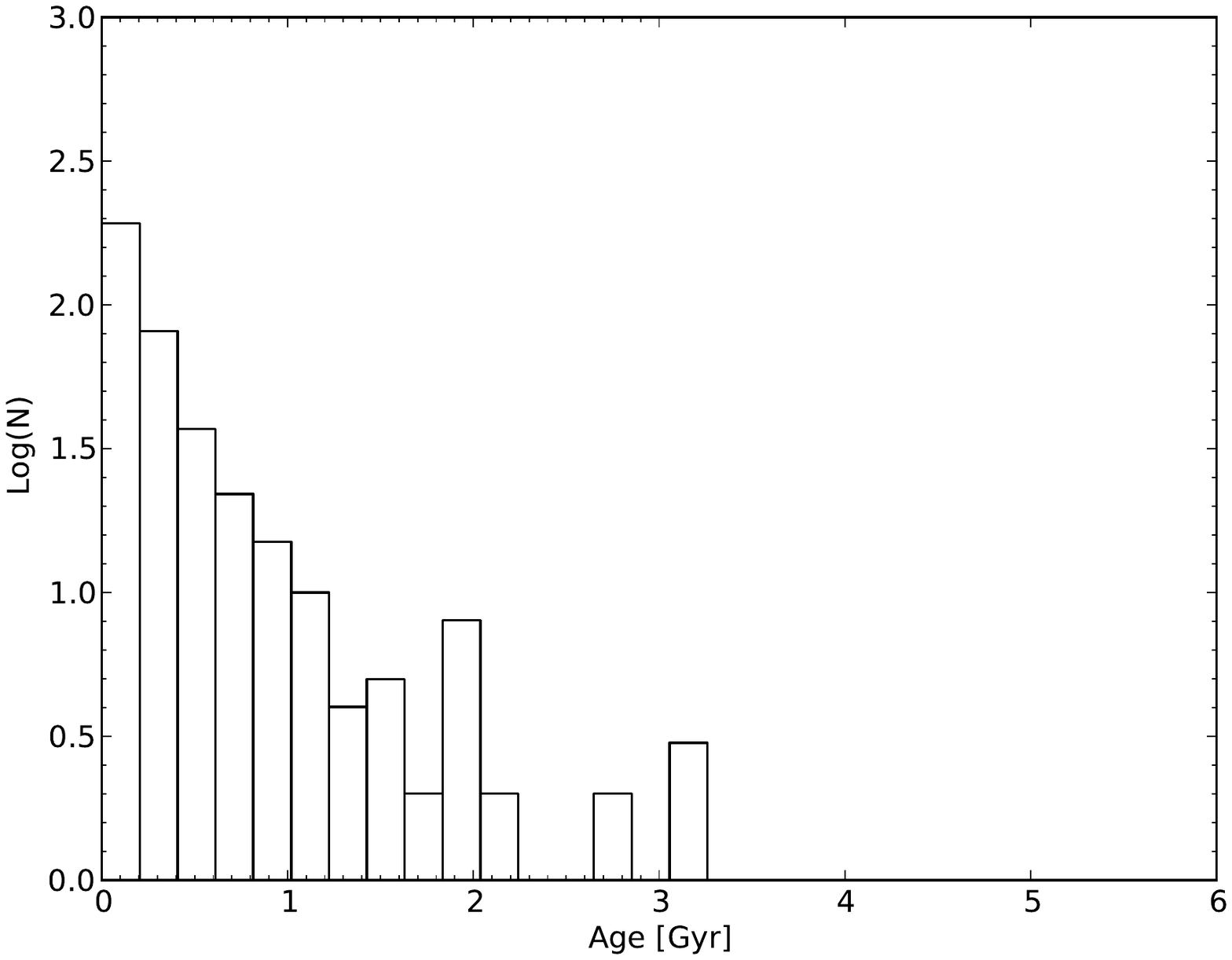} \\
\includegraphics[width=8.6cm,angle=0]{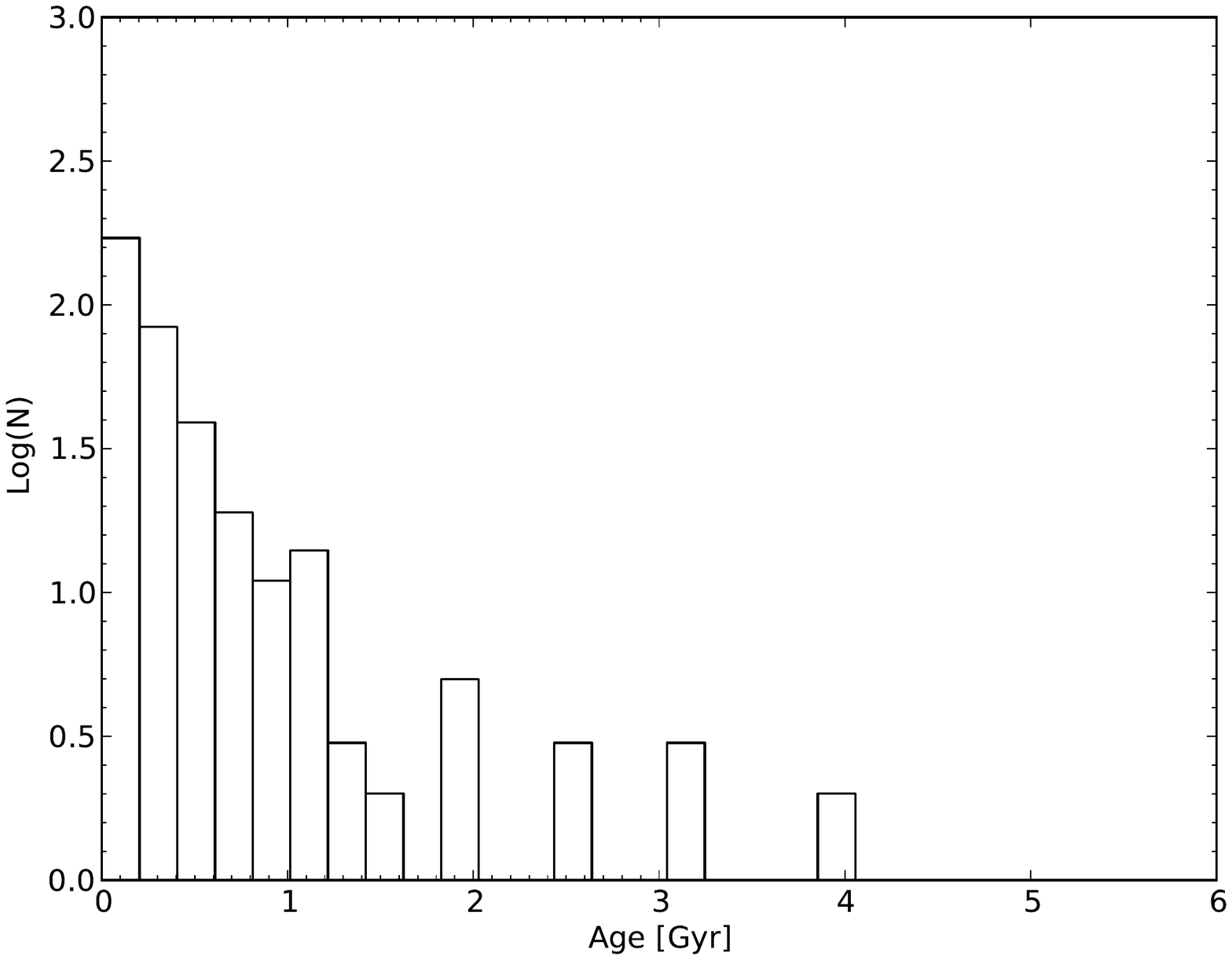} \hfill
\includegraphics[width=8.6cm,angle=0]{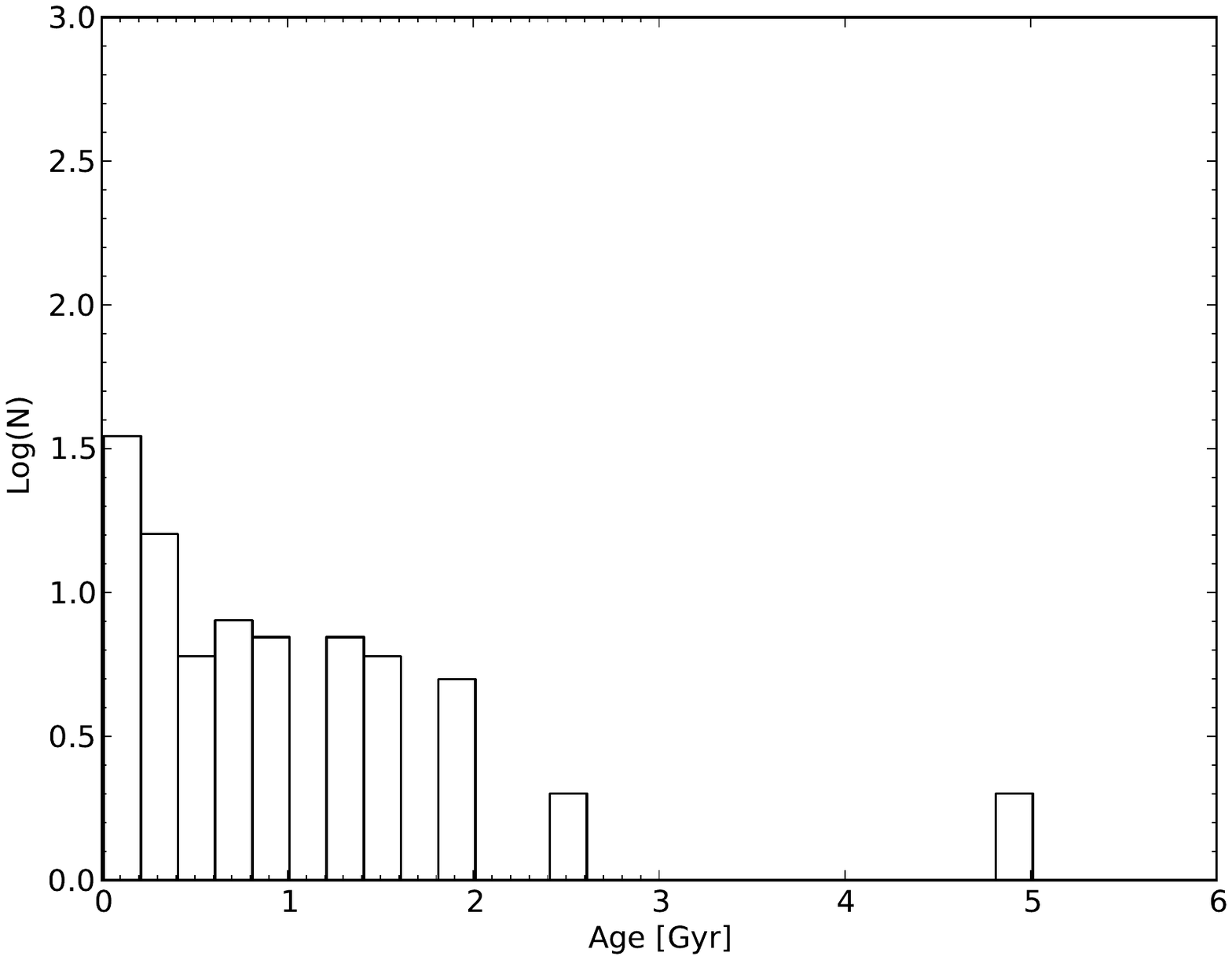}

\caption{\label{agedistributions}  Age distribution of the homogeneously
selected cluster sub-samples used in our work (top-left: CS\,1 -- Kharchenko;
top-right: CS\,2 -- Dias; bottom-left: CS\,3 -- WEBDA; bottom-right: CS\,4 --
FSR).  }.

\end{figure*}

\section{Isochrone Fitting}\label{sect_age}

For all of the above mentioned cluster samples, except the FSR clusters, there
are ages available. To perform our analysis we hence need to determine ages for
all the FSR objects. In the following section we detail our approach to fit
isochrones with particular emphasis on performing these fits in an unbiased and
homogeneous way and to obtain accurate ages.

For each FSR object the most likely cluster members are identified using the
established photometric decontamination technique detailed in Paper\,I
(Sect.\,\ref{sect_decon}). We then fit solar metallicity Geneva
\citep{2001A&A...366..538L} or pre-main sequence \citep{2000A&A...358..593S}
isochrones (where appropriate) to the near infrared 2MASS colour-magnitude data
of the highest probability cluster members (Sect.\,\ref{sect_isofit}). As
starting point we utilise our homogeneously determined distance and extinction
values from Paper\,I. All clusters are then fit three times blindly (without
knowledge which cluster is fit) and in a random order. The three  values for
age, distance and reddening are averaged to obtain the final cluster parameters.

\subsection{Cluster Membership Probabilities}\label{sect_decon}

To fit isochrones to NIR colour magnitude diagrams of clusters situated along
the crowded Galactic Plane, the photometry needs to be decontaminated from
foreground and background objects. Otherwise cluster features such as the main
sequence and red giant branch are difficult to identify. We have detailed our
approach to determine membership probabilities for individual stars in each
cluster in Paper\,I. In the following we just provide a short overview of our
method.

The photometric decontamination procedure was originally outlined in
\citet{2007MNRAS.377.1301B} and is based on earlier works by e.g.
\citet{2004A&A...415..571B}. \citet{2010MNRAS.409.1281F} have slightly adapted
the original method to identify cluster members and we have applied the same
procedure in Paper\,I and for the work presented here.

JHK photometry from the 2MASS \citep{2006AJ....131.1163S} point source catalogue
is utilised for all stars in a cluster with a photometry quality flag of
Qflag='AAA'. The radius of the circular cluster area ($A_{cl}$) around the
cluster centre is chosen as one or two times the cluster core radius. The
control area ($A_{con}$) is a ring with an inner radius of five core radii and
an outer radius of $0.5^\circ$. We define the Colour-Colour-Magnitude distance,
$r_{ccm}$, between the star, $i$, and every other star $j \ne i$ in the cluster
area as:

\begin{equation}\label{eq_rccm}
r_{ccm}=\sqrt{\frac{1}{2} \left( J_i - J_j \right)^2 + \left( JK_i - JK_j
\right)^2 + \left( JH_i - JH_j \right)^2},
\end{equation}

where $JK = J-K$ and $JH = J-H$ are the 2MASS NIR colours. We then determine
$r^{N}_{ccm}$ as the distance to the $N^{\rm th}$ nearest neighbour to star $i$
within the cluster area in this Colour-Colour-Magnitude space. As detailed in
Paper\,I, the exact choice of the value for $N$ will not influence the results,
i.e. the identification of the most likely cluster members. Thus, in accordance
to our procedure in Paper\,I we set $N=25$. We then count the number of stars
($N^{con}_{ccm}$) in the control field that are closer to star $i$ in the
Colour-Colour-Magnitude space than $r^{N}_{ccm}$. Normalising this number by the
respective area allows us to determine the membership-likelihood index or
cluster membership probability ($P^i_{cl}$) of star $i$ via:

\begin{equation}\label{eq_pcl}
 P^i_{cl}=1.0-\frac{N^{con}_{ccm}}{N}\frac{A_{cl}}{A_{con}}.
\end{equation}

Should statistical fluctuations lead to negative $P^i_{cl}$ values, then the
membership probabilities for this particular star are set to zero. Note that we
are only interested in the most likely cluster members, whose $P^i_{cl}$ values
will not be influenced by this.

\subsection{Isochrone fitting}\label{sect_isofit}

\begin{figure*}
\includegraphics[width=8.6cm,angle=0]{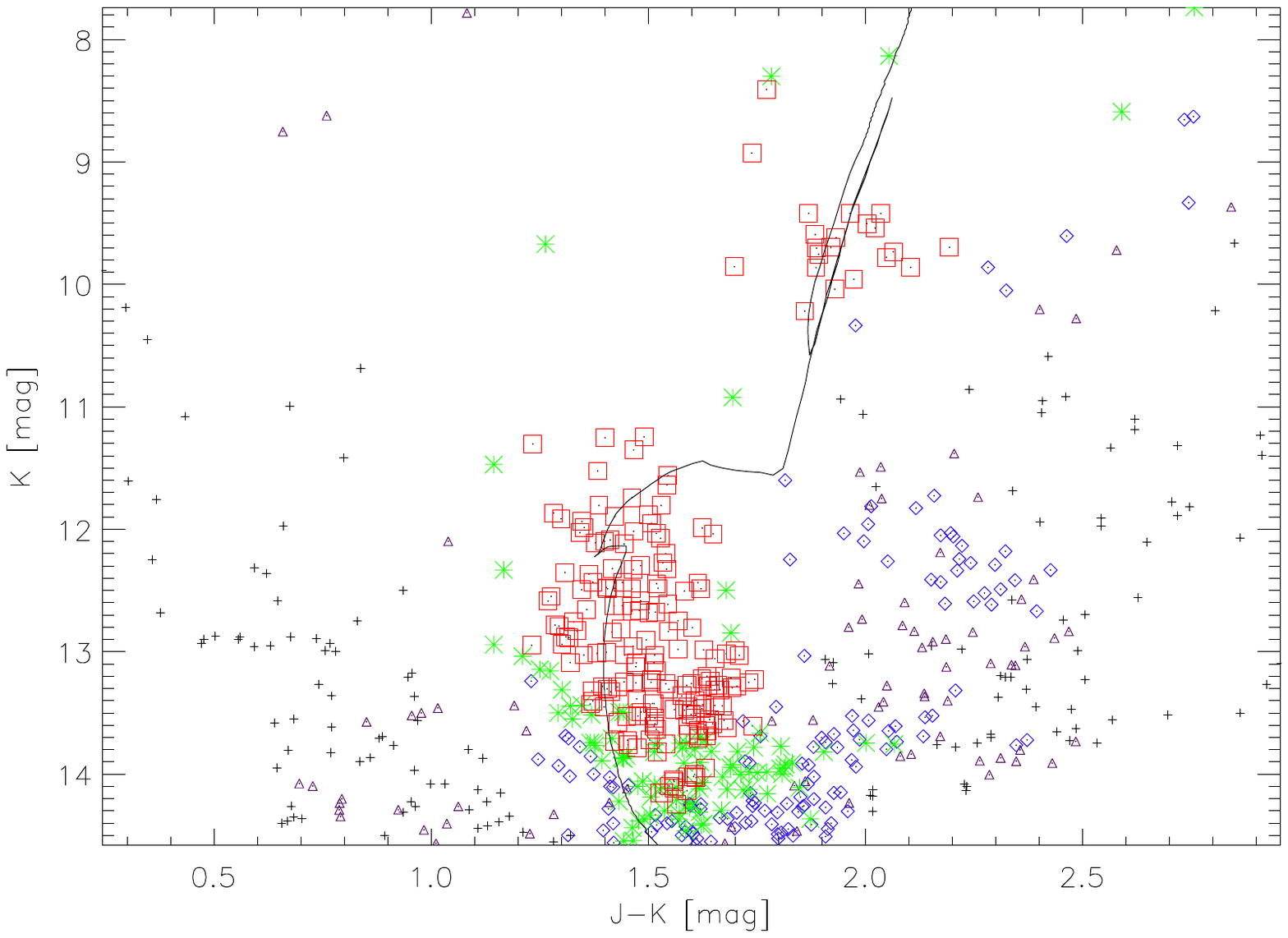} \hfill
\includegraphics[width=8.6cm,angle=0]{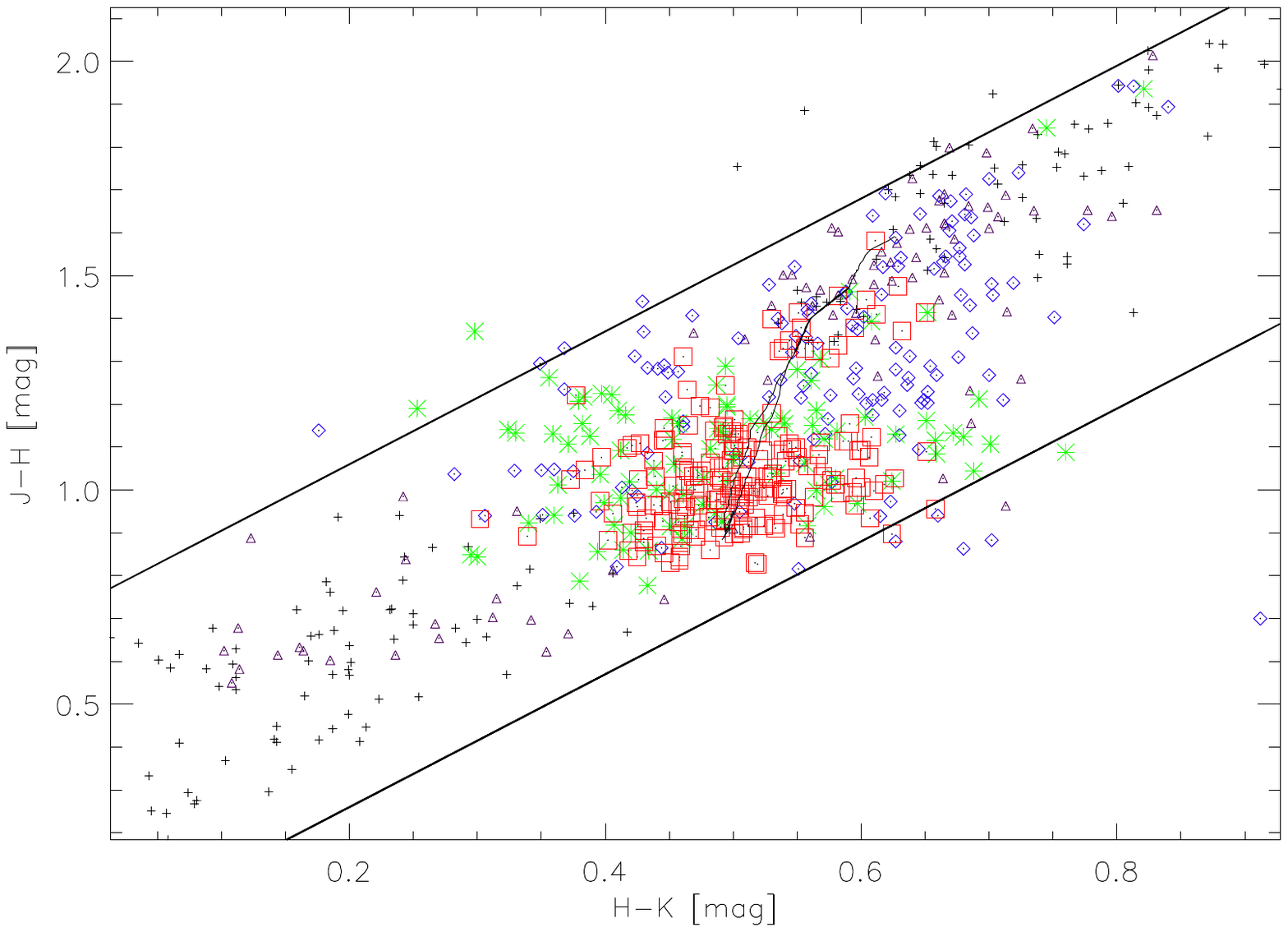}

\caption{\label{fig_iso} Isochrone fits for all stars within two cluster core
radii for FSR\,0233. Symbols represent the determined cluster membership
probabilities:$P^i_{cl} > 80$\,\% red squares; $60\% <P^i_{cl}< 80$\,\% green
stars; $40\% < P^i_{cl} < 60$\,\% blue diamonds; $20\%<P^i_{cl}< 40$\,\% purple
triangles; $P^i_{cl}< 20$\,\% black plus signs. The left panel, shows the
isochrone fit in the $J-K/K$ colour magnitude space, the right panel shows the
isochrone fit in the $H-K/J-H$ colour colour space. The overplotted isochrone
(black solid line) has the parameters of $log(age/yr)=9$, distance of
$d=1.6$\,kpc and H-band extinction $A_H=1.3$\,mag.}

\end{figure*}

Using the above determined cluster membership probabilities for stars in each
cluster region, we utilise NIR colour-magnitude and colour-colour diagrams to
fit isochrones to the data (see Fig.\,\ref{fig_iso} for an example). Since we
have no data available on the metallicities of the clusters, we homogeneously
assume solar metal content. This could be not appropriate for particular
clusters, whose [Fe/H] might range from -0.4 to +0.2, but statistically this
assumption is justified. Furthermore, the median metallicity of all our clusters
that have a WEBDA counterpart is $Z=0.02$ (i.e. solar). We also note that our
statistical errors of the cluster parameters caused by the manual isochrone fits
are typically of the order of, or larger than, the systematic uncertainties
caused by using a slightly erroneous metallicity. Furthmore, the age binsize
used in our analysis in Sect.\,\ref{sec_ageevolh0} is also of the same size or
larger than potential age variations due to variations in the metallicity.
Hence, the clusters in each bin provide a statistically valid representation of
the age.

As model isochrones we utilise the Geneva Isochrones \citep{2001A&A...366..538L}
for intermediate age and old clusters. In some cases the clusters are obviously
very young, i.e. contain Pre-Main Sequence (PMS) stars. For these objects we
utilise the solar metallicity PMS isochrones from \citet{2000A&A...358..593S}
which cover the stellar mass range of ($0.1 M_\odot < M < 7.0 M_\odot$).

\subsubsection{Unbiased isochrone fits}

Our aim is to determine the cluster properties (age, distance, reddening) and
uncertainties for all FSR clusters in a homogeneous way. In order to achieve
this we set up a manual {\it pipeline} which will be described in the following.

We only select FSR cluster candidates for which we have been able to
automatically measure distance and reddening in Paper\,I. These are 771 of the
FSR objects. The remaining clusters and cluster candidates will have an
insufficient number of high probability cluster members, and hence any attempt
to fit an isochrone to these objects will most likely be impossible or result in
very large uncertainties.

In the literature there are many examples of a single cluster having multiple
determined age, distance and reddening values. One such example is FSR\,1716 (as
discussed in the introduction) for which \citet{2010MNRAS.409.1281F} determined
a distance of 7.0\,kpc and $\log(age/yr)$\,=\,9.3, whereas
\citet{2008A&A...491..767B} determined the cluster to be either
0.8\,kpc$/$7\,Gyr or 2.3\,kpc$/$12\,Gyr. A similar case discussed in
Sect.\,\ref{sect_samp} is Stephenson\,2 (or RSGC\,2). This is a young embedded,
red supergiant rich cluster at a distance of about 6\,kpc (e.g.
\citet{2007ApJ...671..781D, 2013MNRAS.436.1116F}), while
\citet{2013A&A...558A..53K} lists a distance of only 1.13\,kpc. Such
inconsistencies can arise from different interpretations of which stars are
potentially giants in the cluster. To account for these possibilities we decided
to fit an isochrone to each cluster three times using a blind fit (the FSR
number or previous fit results are unknown) and a randomised order.

Thus, one of us performed 2313 manual isochrone fits. In every case neither the
FSR number nor the results from previous fits are known. We start each fit with
plotting the NIR colour-magnitude and colour-colour diagrams (as shown in
Fig.\,\ref{fig_iso}) where stars are coded based on their determined cluster
membership probability. Overlayed on these plots are several Geneva isochrones
of different ages ($\log(age/yr)$\,=\,7, 8, 9, 10) using the distances and
extinction values for this cluster from Paper\,I.

The fitter then categorises the cluster in one of three types: i) unable to fit
any kind of isochrone; no feature(s) resembling a star cluster is visible in the
diagrams, hence the cluster is either not real or the object represents an
overdensity that is too low to reliably identify the position of the most likely
cluster members in the colour-magnitude diagrams; ii) cluster age identified as
young; these objects are then fit by a pre-main sequence isochrone; iii) a clear
intermediate age or old open cluster sequence is visible; for these objects  the
closest fit of the four isochrones is chosen and overlayed with a number of
isochrones with steps in  $\log(age/yr)=0.05$. The then closest fit is used as a
starting point to freely vary all three isochrone parameters (age, distance,
reddening) until a satisfactory fit is obtained. A similar procedure is
performed for the pre-main sequence clusters.

\subsubsection{Cluster characterisation and parameters}\label{sect_isopip}

Once the entire sample of cluster candidates has been fitted by the above
described method, i.e. there are three independent fits and classifications for
each candidate, the results for each cluster are combined and objects are
classified into the three categories discussed above.

i) A cluster candidate is considered not a cluster or a too low significant
overdensity if it has been placed at least twice into this category, or if it
has been placed in each of the three categories once.

ii) An object is considered a PMS cluster if it has been placed at least twice
into this category.

iii) An object is considered an open cluster if it has been placed at least
twice into this category.

For the latter two categories we determine the cluster parameters (distance,
age, extinction) as averages from the respective isochrone fits (either three or
two). The resulting values are listed in the Appendix in Table\,\ref{app_table}.
The uncertainties listed in Table \ref{app_table} are then the mean absolute
statistical variations of the individual parameter values for each cluster as
obtained by the fitter. Note that they do not include any systematic
uncertainties caused by using solar metallicity isochrones.

\section{Scale Height determination}

\subsection{Cluster distribution functions}

In order to analyse the distribution of star clusters perpendicular to the
Galactic Plane, one can assume that the space density $N(Z)$ of clusters as a
function of the height $Z$ above/below the plane follows a certain analytical
function. This could be for example an exponential distribution of the form

\begin{equation}\label{eq_fz1}
N(Z)=N_{0} \cdot \exp \left(- \frac{\left| Z-Z_{0} \right| }{h_0} \right),
\end{equation}

or

\begin{equation}\label{eq_fz2}
N(Z)=N_{0} \cdot sech^2 \left( \frac{ \left| Z-Z_{0} \right| }{2 \cdot h_0} \right),
\end{equation}

which is to be expected for a self-gravitating disk. In both equations $N_0$
gives the central space density of clusters at $Z=Z_0$, where $Z_0$ is the
vertical centre (zero point) of the distribution and $h_0$ is the scale height.
Both distributions are very similar within a few scale heights, and are in fact
identical at $\left|Z-Z_0\right| = h_0$.

We plan to investigate the evolution of the scale height $h_0$ as a function of
cluster age and also the distance of the clusters from the Galactic Centre. The
cluster samples we can utilise usually only include objects at most a few scale
heights from the mid-plane. It is hence not of relevance which parametrisation
we utilise and we chose the exponential distribution for the purpose of this
paper.

Furthermore, our sample sizes to determine the free parameters of this
distribution ($N_0, Z_0, h_0$) are going to be small. Hence, any algorithm to
determine these parameters needs to be robust for small samples and also allow
us to estimate realistic uncertainties for each of the parameters in order to
reliably infer trends or to identify differences in e.g. $h_0$ which are
statistically significant. Note that a simple exponential fit to a histogram for
the $Z$ distribution of clusters is not sufficient for this purpose, as it will
break down easily even for sample sizes of the order of 100 clusters (see e.g.
\citet{2006A&A...446..121B}, \citet{2006A&A...445..545P}).

\subsection{Parameter determination}\label{sect_sh}

In order to ensure reliable values for the parameters ($N_0, Z_0, h_0$) and
accurate uncertainties even for small cluster samples ($N < 100$), we compare
the distribution of $Z$-values of our sample with model distributions via a two
sample Kolmogorow-Smirnov (KS) test \citep{1983MNRAS.202..615P}. The model
distributions are obtained for different scale heights and $Z_0$ values. The
parameters of our cluster sample are taken as the values of the model
distribution which shows the highest probability to be drawn from same parent
distribution.

\subsubsection*{Model Distribution Size}

The 2-sample KS-test uses a Cumulative Distribution Function (CDF) for the two
samples of $Z$ values to estimate the probability $P_{KS}$ that both are drawn
from the same parent distribution. Our model sample of clusters will have to
have at least the same range of $Z$-values as the observed sample whose
parameters we are trying to determine. With the known $Z_{min}$ and $Z_{max}$
values of the observed sample, in principle we can determine an analytical
expression for the CDF of the model by integrating Eq.\,\ref{eq_fz1} along $Z$.
However, we decided to obtain this CDF by generating a sample of $N_M$
$Z$-values randomly distributed according to Eq.\,\ref{eq_fz1}.

The size of $N_M$ should be as small as possible to limit the computing time,
but as large as required to remove any uncertainties due to the random nature of
the sample. We hence determined $P_{KS}$ values of an observed cluster sample
against model cluster samples with $N_M$ $Z$-values. The size $N_M$ of the model
cluster sample was varied from 300 to 50.000 objects. For each $N_M$-value we
repeated these tests multiple times with different random realisations of the
distribution of $Z$-values. The size $N_M$ of the model sample was judged to be
sufficient when for 9 out of 10 random realisations the $P_{KS}$ were identical.
This occurred for model sizes of about $N_M = 30,000$. Note that we have
repeated these tests for multiple combinations of $h_0$ and $Z_0$ values in the
model, with no changes to the results. Hence, all our model cluster samples
contain 30,000 clusters.

\subsubsection*{Model Parameter Ranges}

As mentioned above, all our model distributions will contain $Z$-values for
30,000 objects within the minimum and maximum $Z$-value of the observed
distribution whose parameters we are trying to determine. We want to determine
the parameters ($h_0, Z_0$) of the observed distribution without any prior
assumptions. Thus, we generated model distributions where the parameters $h_0$
and $Z_0$ did span the entire possible parameter space. In other words we varied
$h_0$ between 20\,pc and 1000\,pc, while $Z_0$ had values between -160\,pc and
+100\,pc. In both cases 5\,pc increments where chosen for both parameters. This
resulted in 197\,x\,53\,=\,10,441 different model distributions for each of the
observed cluster samples.

\begin{figure}
\includegraphics[width=8.6cm,angle=0]{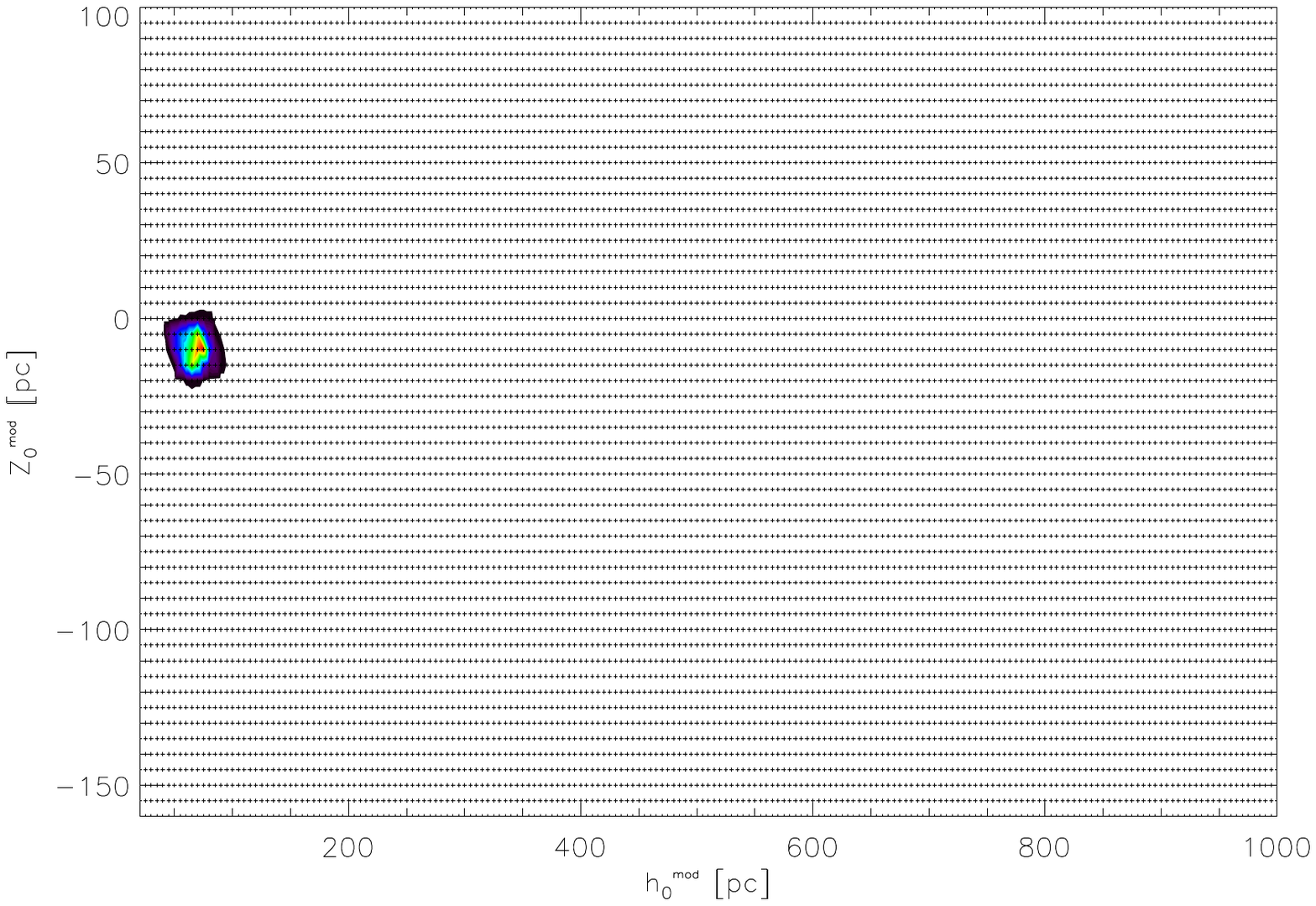}

\caption{\label{fig_contour_grid} Plot of the $P_{KS}$ values for an observed
cluster distribution for the entire modelled $h_0$ vs $Z_0$ parameter space.
Crosses indicate the positions for which we performed a KS-test. The
colours/contours indicate the probabilities that the modelled and observed
distributions are drawn from the same parent sample. Most of the $P_{KS}$ values
are almost zero (white, lowest contours), and the highest non-zero values (red,
highest contours) are only found in a small area of the parameter space. The
sample contains all clusters from the MWSC catalogue (CS\,1) within our chosen
distance range in the 4th Galactic quadrant. There are 313 clusters in this
sample and we find a best fit for the scale height of 68.1\,pc and the vertical
zero point of -9.9\,pc.}

\end{figure}

\begin{figure*}
\includegraphics[width=8.6cm,angle=0]{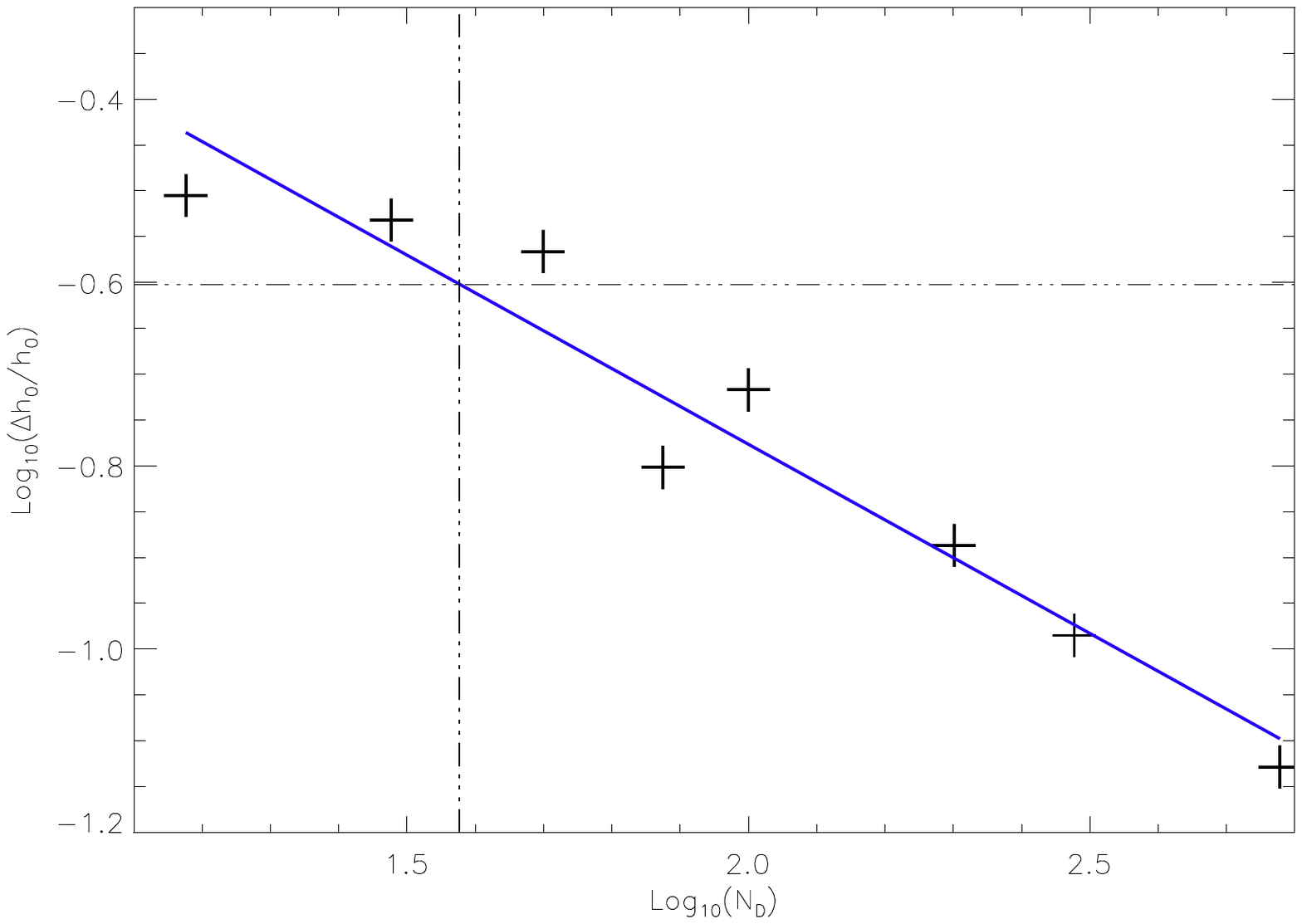} \hfill
\includegraphics[width=8.6cm,angle=0]{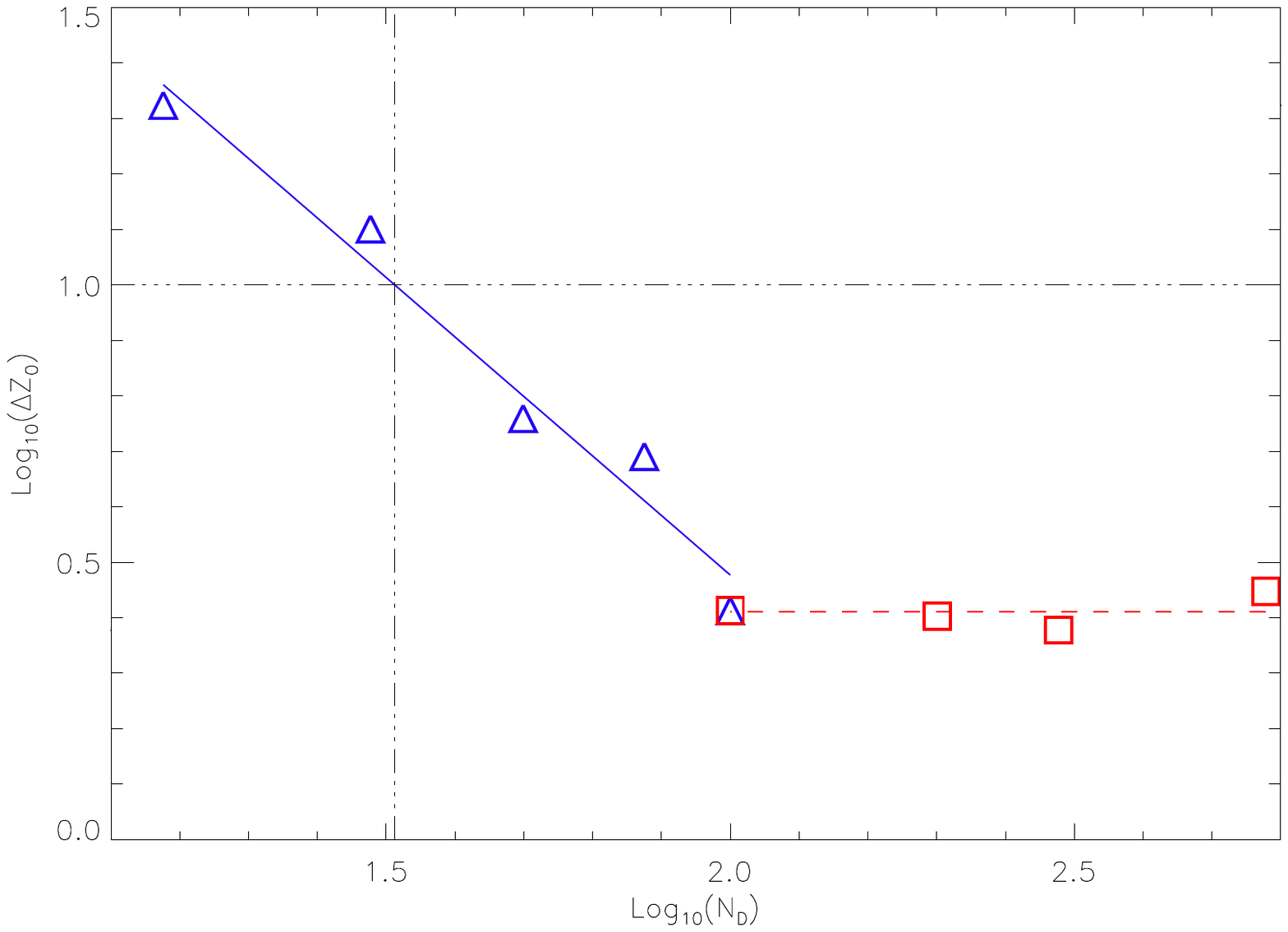}

\caption{\label{z0h_errors} \textit{Left:} Plot of scale height relative error
against sample size. Black crosses represent the mean values for the modelled
distributions. The solid line represents our fitted power law, and the dot-dash
lines a 25\,\% error on the scale height, which is achieved for a sample size of
38 clusters or above. \textit{Right:} Plot of vertical zero point absolute error
against sample size. Blue triangles and red squares represent the mean values
obtained for the modelled distribution. The blue solid line and red dashed line
represent the respective linear fits to sample sizes below and above 100
clusters. The dot-dash lines identifies an error of 10\,pc for $Z_0$ which is
achieved for a sample size of 32 clusters or larger.}

\end{figure*}

\subsubsection*{Best Fit Parameters}

We now perform a 2-sample KS-test of the observed sample against all the 10,441
model distributions to determine the probabilities $P_{KS}$ that the two samples
are drawn from the same parent distribution. In Fig.\,\ref{fig_contour_grid} we
show the distribution of $P_{KS}$-values for one example of an observed cluster
distribution (all selected clusters from CS\,1 in the 4th Galactic Quadrant)
over the entire modelled $h_0$--$Z_0$ parameter space, i.e. the figure shows
$P_{KS}(h_0, Z_0)$. As one can see, for vast regions of the parameter space, the
$P_{KS}$-values are almost zero. Only for a limited area the the values are
non-zero.

In order to find the best fitting parameters for the observed distribution we do
not chose the set of parameters that leads to the highest probabilities
$P_{KS}$. Instead we fit a 2-dimensional Gaussian distribution to the
$P_{KS}(h_0, Z_0)$ values, where the centre and width are free parameters. The
central coordinates of this Gaussian are then taken as the best fit parameters
for the observed distribution.

\subsection{Parameter Uncertainties}\label{sect_accuracy}

Our above described approach generates two best-fit parameters for each observed
cluster distribution. Since we plan to investigate potential changes with age or
Galactocentric distance of the scale height of our observed cluster
distributions, we require to know the uncertainties of our method in order to
judge if any trends in the data are significant. In other words we need to
estimate how large the uncertainties $\Delta h_0$ and $\Delta Z_0$ are and
if/how these uncertainties depend on the value of the parameters and the size of
the cluster sample.

In order to estimate these uncertainties we simulated $Z$-distributions for
small cluster samples with various $h_0$ and $Z_0$ values and processed them
with our above described procedure to determine their scale height and vertical
zero point. Since we know the input parameters for each simulated distribution,
we can evaluate the uncertainty for both parameters by repeating the process
with 50 different random realisations of the simulated $Z$-distributions. The
uncertainties $\Delta h_0$ and $\Delta Z_0$ are estimated as the $rms$ of the
individual measurements $h_{0,i}$ and $Z_{0,i}$ compared to the input values.

To test any dependencies of the uncertainties on the parameter values of $h_0$
and $Z_0$ we did two tests: i) we kept $Z_0$\,=\,-30\,pc and varied $h_0$
between 100\,pc and 350\,pc, which covers the potential range of scale heights
for most of our observed samples; ii) we fixed the scale height to
$h_0$\,=\,200\,pc and varied the vertical zero point of the distribution from
-40\,pc to +40\,pc. In both cases no significant or systematic dependence of the
uncertainties on the parameter values is found.

More importantly, we also need to test \textit{how} the uncertainties depend on
the sample size $N_D$. We hence repeated all the above tests for simulated
cluster samples with $N_D$\,=\,15, 30, 50, 75, 100, 200, 300 and 600 clusters.
We find that the sample size $N_D$ has a systematic and significant influence on
the uncertainties of both parameters $h_0$ and $Z_0$. In particular we find that
the relative uncertainty of the scale height scales with the sample size $N_D$
approximately as a power law. Also the absolute uncertainty of the vertical zero
point of the distribution scales as an approximate powerlaw with the sample
size, but only for small samples. Above a sample size of about 100 clusters, the
absolute uncertainty of $Z_0$ remains constant. This is shown in
Fig.\,\ref{z0h_errors}.

From our powerlaw fits we can hence calculate the uncertainties from our method
solely from the knowledge of the sample size $N_D$ using the following
equations:

\begin{equation}\label{eq_herr}
\frac{\Delta h_0}{h_0}=1.12 \cdot \left( N_{D} \right)^{-0.41}
\end{equation}

\begin{equation}\label{eq_zerr}
\Delta Z_0 = \left\{  \begin{array}{l l} 419\,\text{pc} \cdot \left(
N_D\right) ^{-1.07}  & \quad \text{if $N_D < 115$}\\ 2.6\,\text{pc} & \quad
\text{if $N_D > 115$} \end{array} \right.
\end{equation}  

In other words, the relative uncertainty of the scale height scales roughly with
the inverse of the square root of the sample size, while the absolute uncertainty
of the zero point scales as the inverse of the sample size. We believe that the
constant uncertainty of the zero point $Z_0$ above a sample size of about 100
clusters is caused by our step size of 5\,pc in the model distributions. For
such large samples the uncertainty becomes smaller than half of our step size,
which then becomes the limiting factor compared to the sample size. Should
higher accuracies for $Z_0$ be required, the step size can be decreased. We
refrain from this in this paper, since we judge 2.6\,pc as uncertainty for $Z_0$
for large samples sufficient.

\section{Results and Discussion}\label{sect_results}

\subsection{FSR cluster characterisation and parameters}\label{results_pipeline}

Our data-processing pipeline was applied to the sub-sample of 775 FSR List
clusters which had a distance and extinction values determined in Paper\,I. Here
we successfully determine the ages of 298 clusters. All their parameters and
respective uncertainties are listed in the Appendix in Table\,\ref{app_table}.
Hence, only about 40\,\% of the investigated FSR clusters passed our stringent
criteria for a successful isochrone fit. Of those, 216 are flagged as previously
'known', and 82 as 'new' in the FSR catalogue. Note, that 'new' stands for
clusters that are new discoveries in \citep{2007MNRAS.374..399F}. Thus, we
confirm here that these 82 previously unknown objects are in fact real clusters
and determine their parameters.

The low percentage of these 'new' clusters in the entire sample can be
interpreted in two ways: (i) A large fraction of these clusters are
overdensities but not in fact real clusters, i.e. no isochrone could be fitted;
(ii) It is significantly more difficult to fit isochrones to these clusters
since they are less significant overdensities.

\citet{2007MNRAS.374..399F} showed that about half of the entire FSR list of
'new' objects might in fact be not real clusters but overdensities, which was
confirmed through spatial analysis by \citet{2008MNRAS.385..349B} and
\citet{2010A&A...521A..42C}. However, as discussed in Paper\,I, the
contamination of the cluster sub-sample of 775 objects used here is less than
25\,\%, thus at least 75\,\% of the clusters are potentially real.  During the
isochrone fits for the clusters in our FSR sub-sample, it was noted that a large
proportion of clusters had a poorly defined main sequence; in many cases only
the top was visible within the 2MASS magnitude limit and thus an isochrone fit
was not possible under the constraints of our data-processing pipeline. On
completion of the pipeline, we found that a large proportion of the known
objects had a clear and well defined main sequence and$/$or red giants, whereas
the unknown objects had fewer members (hence they remained undetected) whose
main sequences were not as well defined, and in many cases fell below the
magnitude limit of 2MASS. We would argue, therefore, that the low number of
confirmed new clusters in our sample is a reflection of the difficulty involved
in fitting isochrones to the new objects, rather than the majority being
over-densities.
 
We make a comparison of the distance and H-band reddening values determined in
Paper\,I using our novel photometric method ($D^{PI}, A_{H}^{PI}$), and those
from our data-processing pipeline described in Sect \ref{sect_isopip} of this
paper ($D^{P2}, A_{H}^{P2}$). The two distance values depend linearly on one
another, with $D^{P1}\approx 25\%$ larger than $D^{P2}$, with a scatter of
65\,\% and Pearson correlation coefficient of 0.89. The primary source of the
large scatter are clusters concentrated at small distances, i.e. $D^{P2} \le
3kpc$. The scatter decreases with increasing $D^{P2}$. This can be explained
since our photometric distance measurement method in Paper\,I works by measuring
the density of stars foreground to a cluster which is more accurate for larger,
more extincted objects.

The two reddening values also depend linearly on one another, agreeing within
$5\%$ with a scatter of $9\%$ and Pearson correlation coefficient of $0.95$.
Unlike $D^{P1}$, the determination $A_{H}^{P1}$ depends only on the ability
to accurately determine a clusters median colour, and hence is independent of
individual cluster reddening values.

Furthermore, we have compared our ages to the ages in MWSC, for the clusters
which are in both lists. There are a few obvious outliers, where ages differ by
a factor of 10 or more. However, after removing those, both ages show a
correlation coefficient of 0.73, with a rms scatter of 0.19 for log(age/yr). The
latter can be interpreted as a more realistic uncertainty of the ages determined
for the FSR clusters, compared to the pure statistical estimates quoted in
Table\,\ref{app_table}.

As already stated in Sect.\,\ref{sect_samp}, the resulting FSR sub sample after
the age determination is only very small. If we further require a homogeneous
completeness for the scale height analysis, the sample size becomes even
smaller. Hence we have not included the FSR-subsample in the scale height
analysis performed in the remainder of the paper. However, as is evident in the
radial distribution (lower right panel of Fig.\,\ref{samples}) and the age
distribution (lower tight panel of Fig.\,\ref{agedistributions}) sample is
dominated by rather old, and distant clusters. They are hence in itself an
important addition to the existing large cluster samples, potentially enlarging
their current radius of completeness.

\subsection{Cluster Scale Height and Zero Point}

Our novel method is designed to determine a cluster sample's scale height
$h_0$ and zero point $Z_0$, whilst significantly reducing the restraint on
sample size. The approach to utilise modelled distributions in conjunction with
KS-tests allows us to determine $h_0$ with a better than 25\,\% accuracy for a
sample of 38 clusters or larger. For the same sample size we can determine $Z_0$
within 8.5\,pc. In the following we hence investigate sub-samples of CS\,1, 2, 3
with roughly this size, in order to establish if there are systematic and/or
significant evolutionary or positional trends in the cluster distribution within
the plane of the Galaxy. We investigate each of the three cluster samples to
find out if there are differences between them that might be caused by potential
biases in the samples.

\begin{figure}
\includegraphics[width=8.6cm,angle=0]{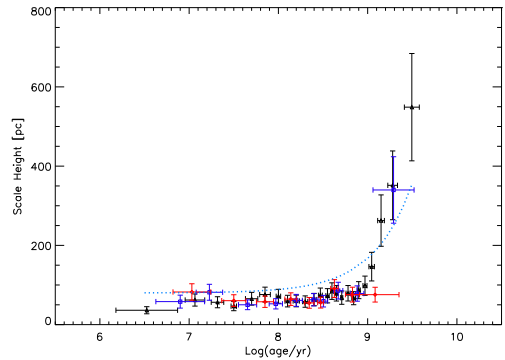}

\caption{\label{fig_h0age} Evolution of the cluster scale height $h_0$ with age
for the 3 investigated samples. Black triangles indicate CS\,1 (MWSC), blue
squares indicate CS\,2 (DAML02) and red diamonds CS\,3 (WEBDA). The horizontal
'age' error bars indicate the typical rms of log(age/yr) from the median age in
each bin. The dashed line is the approximate scale height -- age relation for
field stars (see text for details).}

\end{figure}

\subsubsection{$h_0$ Evolution with Age}\label{sec_ageevolh0}

We investigate how scale height changes with cluster age. In Fig \ref{fig_h0age}
we show the scale height values we derived using our method over a range of age
bins. The age ranges for each bin and the number of clusters in them for every
CS are listed in Table\,\ref{age_bins}. There is a general trend of increasing
scale height with cluster age. Most notably there is an apparent marked increase
in the gradient at $log(age/yr)  = 9$ or a cluster age of about 1\,Gyr. We
perform a linear fit of the scale height against $log(age/yr)$ and find that the
observational trend in Fig.\,\ref{fig_h0age} can be characterised by:

\begin{equation}\label{eq_h0age}
h_0 \propto \left\{  \begin{array}{rl} 11.0\,\text{pc} \cdot log(age/yr) & \quad
\text{if age $\le$ 1\,Gyr}\\ 880\,\text{pc} \cdot log(age/yr)  & \quad \text{if
age $\ge$ 1\,Gyr} \end{array} \right.
\end{equation}
where $h_0$ is the scale height and $log(age/yr)$ is the cluster age. Note that at an
age of 10\,Myr, the time when gas expulsion has typically finished, the scale
height of the clusters is about 50\,pc. Please note that the above given values
for the changes of scale height with cluster age are independent of the actual
choice of the borders for our age bins. The only sample where the marked change
in behaviour at 1\,Gyr is not evident is CS\,3 -- WEBDA. The reason is that in
our homogeneously selected sub-sample there are simply not enough old clusters to
trace $h_0$. In particular the oldest age bin spans a factor of 14 in age (see
Table\,\ref{age_bins}), but is dominated by clusters of an age of 1\,Gyr. CS\,1
and CS\,2 show essentially the same behaviour for older objects (see
Fig.\,\ref{fig_h0age}), even if there is just one 'old' age bin for CS\,2.

\begin{table}

\caption{\label{age_bins} Age bins (minimum and maximum ages) and respective
number of clusters in them for the clusters samples, used in the investigation
of scale height with cluster age. We also list the determined scale height and
zero point with their respective uncertainties.}

\renewcommand{\tabcolsep}{4pt}

\begin{center}
\begin{tabular}{c|cc|c|cc|cc}
\hline
CS &Age$_{\rm min}$ & Age$_{\rm max}$ & N$_{\rm cl}$ &
$Z_0$ & $\Delta Z_0$ & $h_0$ & $\Delta h_0$\\
& \multicolumn{2}{|c|}{[log(age/yr)]} & & \multicolumn{2}{|c|}{[pc]}&
\multicolumn{2}{|c|}{[pc]} \\
\hline
MWSC & 6.000 & 6.850 & 40 & -2 & 8 & 36 & 9 \\
MWSC & 6.850 & 7.200 & 40 & 9 & 8 & 62 & 15 \\
MWSC & 7.200 & 7.420 & 40 & 6 & 8 & 56 & 14 \\
MWSC & 7.420 & 7.550 & 40 & -20 & 8 & 47 & 12 \\
MWSC & 7.550 & 7.755 & 40 & -10 & 8 & 65 & 16 \\
MWSC & 7.760 & 7.950 & 40 & -6 & 8 & 76 & 19 \\
MWSC & 7.950 & 8.060 & 40 & -20 & 8 & 72 & 18 \\
MWSC & 8.060 & 8.150 & 40 & -3 & 8 & 59 & 15 \\
MWSC & 8.150 & 8.255 & 40 & -1 & 8 & 60 & 15 \\
MWSC & 8.255 & 8.350 & 40 & -9 & 8 & 58 & 14 \\
MWSC & 8.350 & 8.445 & 40 & -13 & 8 & 63 & 16 \\
MWSC & 8.445 & 8.505 & 40 & -22 & 8 & 74 & 18 \\
MWSC & 8.505 & 8.580 & 40 & -9 & 8 & 73 & 18 \\
MWSC & 8.585 & 8.632 & 40 & -24 & 8 & 85 & 21 \\
MWSC & 8.635 & 8.690 & 40 & -29 & 8 & 79 & 20 \\
MWSC & 8.695 & 8.735 & 40 & -8 & 8 & 68 & 17 \\
MWSC & 8.735 & 8.800 & 40 & -25 & 8 & 79 & 20 \\
MWSC & 8.800 & 8.865 & 40 & -12 & 8 & 67 & 16 \\
MWSC & 8.870 & 8.930 & 40 & -9 & 8 & 87 & 21 \\
MWSC & 8.935 & 9.005 & 40 & -23 & 8 & 98 & 24 \\
MWSC & 9.005 & 9.100 & 40 & -39 & 8 & 146 & 36 \\
MWSC & 9.100 & 9.200 & 40 & 8 & 8 & 263 & 65 \\
MWSC & 9.200 & 9.400 & 40 & -109 & 8 & 352 & 87 \\
MWSC & 9.400 & 9.700 & 40 & -56 & 8 & 549 & 135 \\ \hline
DAML02 & 6.00 & 7.02 & 29 & -35 & 11 & 58 & 16 \\
DAML02 & 7.03 & 7.50 & 40 & -41 & 8 & 82 & 20\\
DAML02 & 7.50 & 7.83 & 40 & -25 & 8 & 50 & 12\\
DAML02 & 7.84 & 8.09 & 40 & -13 & 8 & 53 & 13\\
DAML02 & 8.09 & 8.30 & 40 & -16 & 8 & 61 & 15\\
DAML02 & 8.30 & 8.45 & 40 & -13 & 8 & 63 & 16\\
DAML02 & 8.45 & 8.60 & 40 & 11 & 8 & 59 & 15\\
DAML02 & 8.60 & 8.78 & 40 & -23 & 8 & 86 & 21\\
DAML02 & 8.78 & 9.01 & 40 & -13 & 8 & 78 & 19\\
DAML02 & 9.03 & 9.90 & 40 & -20 & 8 & 340 & 84 \\ \hline
WEBDA & 6.00 & 7.17 & 38 & -51 & 9 & 82 & 21\\
WEBDA & 7.20 & 7.66 & 40 & -25 & 8 & 60 & 15\\
WEBDA & 7.68 & 8.00 & 40 & -23 & 8 & 58 & 14\\
WEBDA & 8.00 & 8.23 & 40 & -4 & 8 & 64 & 16\\
WEBDA & 8.23 & 8.42 & 40 & -26 & 8 & 55 & 14\\
WEBDA & 8.42 & 8.54 & 40 & 5 & 8 & 56 & 14\\
WEBDA & 8.55 & 8.69 & 40 & -21 & 8 & 91 & 22\\
WEBDA & 8.69 & 8.95 & 40 & -34 & 8 & 76 & 19\\
WEBDA & 8.96 & 10.12 & 40 & -9 & 8 & 76 & 19\\
\hline
\end{tabular}
\end{center}
\end{table}

Previous efforts to determine the $h_0$ of older clusters as a function of age
have had limited success. Restrictions on sample size caused by the small size
of the older cluster sample and the spread of their distributions with
increasing age, has until now prevented a detailed analysis of evolution of the
scale height of old clusters. Attempts to place a value on the scale height have
yielded a value of  $h_0=375 pc$ for clusters older than 1\,Gyr (e.g.
\citet{2010MNRAS.409.1281F}). From Eq.\,\ref{eq_h0age} and Fig. \ref{fig_h0age},
this value corresponds to an age of about 2.2\,Gyr i.e. in the middle of the
'old' cluster age bin. Hence this literature value is an average scale height
for clusters older than 1\,Gyr.  Figure\,\ref{fig_h0age} also demonstrates the
superiority of the MWSC list in combination with our novel approach to determine
the scale height, as the larger sample size of CS\,1 allows us to clearly trace
the scale height evolution for objects older than 1\,Gyr in several bins and to
show that there is a systematic significant observational trend in the cluster
scale height with age for objects up to 5\,Gyr.

To the best of our knowledge there are currently no numerical investigations of
the scale height of stellar clusters as a function of age in the Galactic Plane.
This is most likely due to the complexity of the problem which requires 
following the evolution of individual stars in clusters of varying mass to
account for the cluster dissolution over time, as well as the cluster as a whole
in the gravitational potential of the Galactic Disk. However, we can try to
compare the scale heights of objects of different ages with the here determined
evolution of $h_0$ for clusters to infer the basic physical reasons for the
evolution, and in particular the marked change in behaviour after about 1\,Gyr.

The dust in the Galactic Plane has a scale height of about 125\,pc
\citep{Drimmel2003,Marshall2006} in the vicinity of the Sun. At an age of
1\,Myr, Fig.\,\ref{fig_h0age} and Eq.\,\ref{eq_h0age} show that young star
clusters have a scale height of 40\,pc. This is right in the middle of the range
of scale heights estimated for massive OB-stars (30\,--\,50\,pc;
\citep{Reed2000,Elias2006}). Since the formation of these massive stars is
inextricably linked to clustered star formation, this is expected. Thus, the
formation of massive stars and clusters is only possible within the densest part
of the ISM (within one third of the dust scale height), and no significant
fraction of locally observable clusters (without OB-stars) forms in lower
density environments, further away from the disk midplane.

The number of clusters declines over time (see Fig.\,\ref{agedistributions})
which is well known and understood from numerical models (e.g.
\citet{2008IAUS..246..171G,IAU:7064088,2006A&A...455L..17L,2005A&A...441..117L}).
Causes of disruption timescales depend on both internal and external processes
such as e.g. stellar evolution, tidal stripping and relaxation, shocking by
spiral arms and encounters with giant molecular clouds. A consensus in the
literature has not yet been reached on the role that cluster mass plays in
disruption (for a discussion see e.g. \citet{2011sca..conf...85B}). The dominant
disruption process at a few 100\,Myr is stellar evolution through member loss.
In combination with external processes, clusters may gain enough energy from the
ejection of low mass members to cause the observed moderate changes in scale
height during that phase. We find a 10\,pc increase in $h_0$ per $dex$ in
cluster age from the formation to 1\,Gyr, but the correlation coefficient is
only 0.5, and as low as 0.1 when only considering the first 300\,Myr of
evolution. Thus for the first few 100\,Myr the data suggest no evolution in
$h_0$, but the scale height at an age of about 1\,Gyr reaches about 75\,pc. This
is comparable or smaller than the scale height of other young objects in the
disk (e.g. 130\,pc for bipolar PNs \citep{1995A&A...293..871C}; 55\,--\,120\,pc
for young WDs \citep{2012MNRAS.426..427W})

After the surviving clusters reach an age of about  1\,Gyr, or a scale height
of 75\,pc, there is an apparent sudden increase in $h_0$ corresponding to a
change in the evolutionary behaviour. The increase in scale height is about
880\,pc/$dex$ in age. It has been shown that, assuming mass dependent
disruption, clusters with a mass of less than $10^{4} M_{\odot}$ and within
1\,kpc of the Sun are disrupted after 1\,Gyr \citep{2003MNRAS.338..717B}. Hence,
we expect the cluster sub-samples after 1\,Gyr to be dominated by initially
massive clusters.  Thus, if clusters have survived for this duration, they must
have been scattered into an orbit which places them preferentially far away from
the Galactic mid-plane. This enables them to spend much less time in the denser
parts of the Galactic Disk, decreasing their probability for disruption via
external process (spiral arms, encounters with GMCs) and increasing their
chances of prolonged survival.  In other words, the increase in scale height
also implies that the  population of old clusters is dominated by objects that
have  undergone at least one violent interaction event in their past  that has
moved them into an orbit inclined to the Galactic  Plane. This observational
evidence should hence be able to put  tighter constraints onto comprehensive
numerical models of cluster evolution and disruption in the context of the
entire Galactic Disk.

Note that the behaviour of the scale height for clusters is markedly different
to estimates for field stars. To illustrate this we have overplotted the
principle trend observed for main sequence field stars of varying ages in
Fig.\,\ref{fig_h0age}. This qualitative trend has been obtained by utilising
colour dependent velocity dispersions for main sequence stars presented in
\citet{1998MNRAS.298..387D}. As one can see in Fig.\,\ref{fig_h0age}, the {\it
heating} of the stellar compontent of the disk occurs gradually, while for the
cluster component there is a discontinuity around 1\,Gyr. This demonstrates the
difference of the underlying physical mechanisms for the increase in scale
height. While the stellar component is {\it heated} via N-body interactions, the
tidal field and GMCs, the clusters have a much stronger rate of disappearing
from the observational sample with increasing age, and are only moved to large
scale heights (and thus able to survive) via interactions with massive objects
such as GMCs.

\subsubsection{$h_0$ Dependence on Galactic Position}

\begin{figure} 
\includegraphics[width=8.6cm,angle=0]{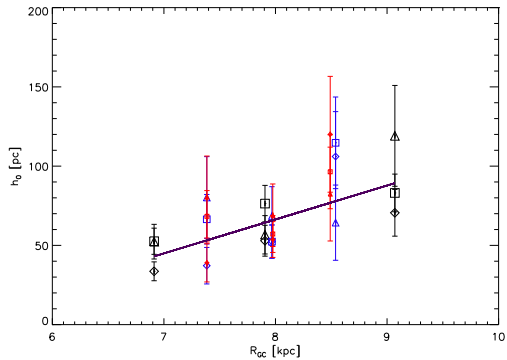}
\caption{\label{fig_h0rgc} Cluster scale height as a function of galactocentric
distance. The symbols indicate different age ranges. Diamonds indicate clusters
younger than 80\,Myr, triangles indicate clusters with ages between 80\,Myr and
200\,Myr and squares indicate clusters with ages between 200\,Myr and 1\,Gyr.
Furthermore, the different CSs are indicated by different colours and symbol
sizes; large black, medium blue, small red for CS\,1, 2, 3, respectively.}
\end{figure}

\begin{table}

\caption{\label{rgc_bins} Age and galactocentric distance ranges used in the
investigation of the dependence of scale height on the galactocentric distance.
We list the cluster sample, age range, $R_{GC}$ range as well as the determined
scale height and zero point with their respective uncertainties. Age bin\,1
corresponds to ages less than 80\,Myr, age bin\,2 corresponds to ages between
80\,Myr and 200\,Myr and age bin\,3 corresponds to ages between 200\,Myr and
1\,Gyr. Older clusters are not included due to the paucity of these objects.}

\renewcommand{\tabcolsep}{4pt}

\begin{center}
\begin{tabular}{c|c|cc|c|cc|cc}
\hline
CS & Age bin & $R_{GC}$ & N$_{\rm cl}$ &
$h_0$ & $\Delta h_0$ & $Z_0$ & $\Delta Z_0$ \\
&  & [kpc] & & \multicolumn{2}{|c|}{[pc]}&
\multicolumn{2}{|c|}{[pc]} \\
\hline
MWSC & 1 & 6.9\,$\pm$\,0.6 & 90 & 34 & 6 & -13 & 3.4 \\
MWSC & 1 & 7.9\,$\pm$\,0.7 & 82 & 53 & 10 & 13 & 3.8 \\
MWSC & 1 & 9.1\,$\pm$\,0.7 & 60 & 71 & 15 & -9 & 5.2 \\
MWSC & 2 & 6.9\,$\pm$\,0.6 & 60 & 52 & 11 & -4 & 5.2 \\
MWSC & 2 & 7.9\,$\pm$\,0.7 & 57 & 57 & 12 & -4 & 5.5 \\
MWSC & 2 & 9.1\,$\pm$\,0.7 & 33 & 119 & 32 & 2 & 9.9 \\
MWSC & 3 & 6.9\,$\pm$\,0.6 & 121 & 53 & 8 & -8 & 2.6 \\
MWSC & 3 & 7.9\,$\pm$\,0.7 & 134 & 76 & 11 & -17 & 2.6 \\
MWSC & 3 & 9.1\,$\pm$\,0.7 & 151 & 83 & 12 & -23 & 2.6 \\ \hline
DAML02 & 1 & 7.4\,$\pm$\,0.4 & 23 & 37 & 12 & -37 & 15 \\
DAML02 & 1 & 8.0\,$\pm$\,0.3 & 66 & 52 & 11 & -13 & 4.7 \\
DAML02 & 1 & 8.6\,$\pm$\,0.3 & 33 & 106 & 28 & -49 & 9.9 \\
DAML02 & 2 & 7.4\,$\pm$\,0.4 & 21 & 80 & 26 & -4 & 16 \\
DAML02 & 2 & 8.0\,$\pm$\,0.3 & 32 & 69 & 19 & -21 & 10 \\
DAML02 & 2 & 8.5\,$\pm$\,0.3 & 15 & 64 & 24 & -26 & 23 \\
DAML02 & 3 & 7.4\,$\pm$\,0.4 & 45 & 66 & 16 & -14 & 7.1 \\
DAML02 & 3 & 8.0\,$\pm$\,0.3 & 68 & 52 & 10 & -4 & 4.6 \\
DAML02 & 3 & 8.5\,$\pm$\,0.3 & 38 & 115 & 29 & -17 & 8.5 \\ \hline
WEBDA & 1 & 7.4\,$\pm$\,0.4 & 23 & 39 & 12 & -41 & 15 \\
WEBDA & 1 & 8.0\,$\pm$\,0.3 & 59 & 54 & 11 & -21 & 5.3 \\
WEBDA & 1 & 8.5\,$\pm$\,0.3 & 24 & 120 & 37 & -72 & 14 \\
WEBDA & 2 & 7.4\,$\pm$\,0.4 & 21 & 80 & 26 & -5 & 16 \\
WEBDA & 2 & 8.0\,$\pm$\,0.3 & 28 & 69 & 20 & -13 & 12 \\
WEBDA & 2 & 8.5\,$\pm$\,0.3 & 16 & 82 & 30 & -36 & 22 \\
WEBDA & 3 & 7.4\,$\pm$\,0.4 & 44 & 68 & 16 & -14 & 7.3 \\
WEBDA & 3 & 8.0\,$\pm$\,0.3 & 64 & 57 & 12 & -17 & 4.9 \\
WEBDA & 3 & 8.5\,$\pm$\,0.3 & 42 & 96 & 23 & -10 & 7.7 \\
\hline
\end{tabular}
\end{center}
\end{table}

We investigate if the cluster scale height changes with Galactocentric radius,
$R_{GC}$. To eliminate the apparent age effects discussed in
Sect.\,\ref{sec_ageevolh0}, we determine $h_0$ for 4 age bins. These are: bin\,1
-- age less than 80\,Myr; bin\,2 -- age between 80\,Myr and 200\,Myr; bin\,3 --
age between 200\,Myr and 1\,Gyr; bin\,4 -- age above 1\,Gyr. Each of these age
bins is separated into 3 ranges for the $R_{GC}$ values per cluster sample. See
Table\,\ref{rgc_bins} for details of each bin. Note that this table does not
contain the details for the oldest age bin\,4, as the paucity of old clusters
did not allow to split them into several $R_{GC}$ bins and still being able to
determine scale height and zero point with sufficient accuracy to draw any
meaningful conclusions. In Fig.\,\ref{fig_h0rgc} we show that there is a
positive trend between $h_0$ and $R_{GC}$ for clusters younger than 1\,Gyr,
which can be expressed as:

\begin{equation}\label{eq_h0rgc}
h_0 [pc] \propto 0.02 \cdot R_{GC}[pc]
\end{equation}

where $h_0$ is scale height and $R_{gc}$ is the median Galactocentric distance
of a cluster sample. There is considerable scatter, but the Pearson Correlation
Coefficient for the data points, determined including the uncertainties, ranges
from 0.75 to 0.85 for the age bins 1\,--\,3. It has a value of 0.80 for the
combined sample of all three age bins shown in Fig.\,\ref{fig_h0rgc}. The trend
of increasing scale height with $R_{gc}$ is virtually identical for the age
bins\,2 and 3, and only slightly stronger for the youngest clusters in bin\,1.
Note that at the solar distance to the Galactic Centre (assumed to be 8\,kpc)
the clusters have a scale height of about 65\,pc.

For some of the above not considered $R_{GC}$ bins of the old clusters (age
above 1\,Gyr), we where able to determine the scale height. The values for $h_0$
are dominated by the younger clusters in the age bin, and all scale heights are
between 200\,pc and 400\,pc. However, no correlation of the scale height with
$R_{GC}$ is evident for these older clusters. This is expected from our findings
in the last section, which indicated that the old objects are dominated by
clusters scattered away from the plane in the past.

As for the age evolution of the scale height, there are to the best of our
knowledge no numerical simulations to investigate this. Hence, our data should
proof vital to constrain potential large scale numerical simulations of cluster
evolution in the Galactic Disk.  However, we can try to understand this weak
observed trend to infer its cause. Since we have eliminated the effect of
cluster age, by considering the different age bins, and have found that there is
almost no evolution of $h_0$ for the first few 100\,Myr, any trend in the scale
height of the cluster sample has to be imprinted on it during the formation.
Indeed there seems to be a moderate flaring of the molecular (star forming)
material in the disk (e.g. \citet{1984ApJ...276..182S,1990A&A...230...21W}).
More massive clusters (which include OB-stars) should also form closer to the
mid-plane. These are the objects which are more likely to survive for a given
time. Thus, potentially the observed effect could be caused by the fact that at
smaller $R_{GC}$ values there are more massive clusters formed, originally
closer to the mid-plane, than further out at larger $R_{GC}$. Hence the scale
height is dominated by originally higher mass clusters towards low $R_{GC}$ and
by less massive clusters at higher $R_{GC}$. However, only detailed numerical
simulations of cluster populations in the Milky Way in combination with accurate
cluster mass estimates, both outside the scope of this work, can investigate
this properly. Note that this weak trend could in part also be explained by a
systematic metallicity gradient in the Galactic Disk.

\subsubsection{Vertical Displacement $Z_0$}

We also investigate how $Z_0$ changes with cluster age and $R_{GC}$.  We find
that there is no dependency of $Z_0$ with any of the parameters for our samples.
This is an expected result as the spatial distribution of clusters should follow
the symmetrical distribution function for vertical displacement above/below the
Galactic plane (Eq.\,\ref{eq_fz1}, \ref{eq_fz2}), such that cluster interactions
and disruptions are also symmetrical. Thus, as $h_0$ increases with cluster age,
$Z_0$ will remain constant and only depend on the position of the Sun with
respect o the plane.

We have used this to average all the $Z_0$ values in our samples to obtain the
mean vertical displacement of the Sun with respect to the Galactic Plane based on
the local distribution of stellar clusters. We find a mean value of $Z_0
=-18.5 \pm 1.2 pc$,  and thus $Z_{\odot} =18.5 \pm 1.2 pc$ (which is in
agreement with accepted literature values based on other objects, see e.g.
\citet{2006JRASC.100..146R}, \citet{1995AJ....110.2183H}).
 
\section{Conclusions} \label{sect_conclusion}

We aim to study the temporal and spatial evolution of the scale height of star
clusters in the Galactic Plane.

In a first step we successfully determined ages of 298 clusters from the FSR
list by \citep{2007MNRAS.374..399F} by fitting isochrones. We used our
automatically determined distances and reddening values from
\citet{2013MNRAS.436.1465B} as starting points. Our FSR sub-sample is dominated
by old objects (age $>$\,500\,Myr) with distances between 1.5\,kpc and 2\,kpc.
The distances and extinction values obtained by the isochrone fitting and our
purely automatic method based on NIR photometry \citep{2013MNRAS.436.1465B} show
a good correlation with Pearson Correlation Coefficients of 0.89 and 0.95,
respectively.

We have developed a novel method to determine the scale height and vertical zero
point of cluster distributions using models and Kolmogorow-Smirnov tests. This
significantly lessens the restraint on the sample size and allows us to measure
scale heights with 25\,\% accuracy for cluster samples as small as 38 objects.
At the same time we are able to infer the sample zero point within 8.5\,pc. For
larger samples these errors can be significantly reduced.

To investigate the temporal evolution of cluster scale height we investigated
homogeneously selected sub-samples of star clusters from four large star cluster
catalogues (MWSC \citep{2013A&A...558A..53K}, DAML02
\citep{2002A&A...389..871D}, WEBDA, FSR \citep{2007MNRAS.374..399F}).  The
selected sub-sample of the FSR list is too small to be included in our
subsequent analysis. We find that most of our results are independent of the
cluster catalogue, despite their very different criteria for cluster inclusion
and parameter estimation. As expected, the MWSC catalogue in combination with
our novel scale height determination method, provides the best 'time resolution'
for our investigation.

We find that star clusters are formed (age 1\,Myr) with a scale height of
40\,pc. This is the same as what has been found for OB-stars
\citep{Reed2000,Elias2006}, demonstrating the link of massive and clustered star
formation. For the next 1\,Gyr the scale height of the surviving clusters only
marginally increases by about 10\,pc per $dex$ in age until it reaches about
75\,pc. The data are in agreement with no evolution of $h_0$ for the first few
100\,Myr.

Fom 1\,Gyr onwards the scale height of the surviving cluster population
increases significantly faster with about 880\,pc per $dex$ in age. The reason
for this is most likely that the old cluster sample is dominated by objects
which have been scattered by one or more interactions with Giant Molecular
Clouds into orbits away from the Galactic Plane. Clusters that do not undergo
such a violent event will stay close to the plane, and not survive to ages of
several Gyr. This is markedly different to the behaviour of the stellar
component in the Galactic Disk.

We further find a weak age-independent trend of cluster scale height with
distance from the Galactic Centre. This might be caused by the mass dependence
of the formation of stellar clusters in the disk or a metallicity gradient. No
significant temporal or spatial variations of the zero point of the cluster
distribution have been found. Based on the cluster distribution we estimate that
the Sun has a position of 18.5\,$\pm$\,1.2\,pc above the Galactic Plane, in
agreement with past measurements using different tracers.

A detailed understanding of the here presented observational evidence can
however only be achieved with numerical simulations of the evolution of cluster
samples in the Galactic Disk. Furthermore, more accurate observational cluster
parameters, such as distances from GAIA, larger complete samples of clusters, as
well as accurate mass estimates for them will certainly aid our understanding of
how the dissolution of clusters over time contributes to the stellar content of
the thin and thick disk of the Galaxy.

\section*{Acknowledgements}
 
We would like to thank H.\,Baumgardt for his comments. A.S.M.\,Buckner
acknowledges a Science and Technology Facilities Council studentship and a
University of Kent scholarship. This publication makes use of data products from
the Two Micron All Sky Survey, which is a joint project of the University of
Massachusetts and the Infrared Processing and Analysis Center/California
Institute of Technology, funded by the National Aeronautics and Space
Administration and the National Science Foundation. This research has made use
of the WEBDA database, operated at the Institute for Astronomy of the University
of Vienna. This publication makes use of data products from the Wide-field
Infrared Survey Explorer, which is a joint project of the University of
California, Los Angeles, and the Jet Propulsion Laboratory/California Institute
of Technology, funded by the National Aeronautics and Space Administration.

\bibliographystyle{mn2e}
\bibliography{references}

\begin{appendix}
\onecolumn
\section{FSR Cluster Property Table}\label{appendix1}
\begin{center}
\begin{longtable}[l]{|c|c|c|cc|c|cc|c|cc|cc|}

\caption{\label{app_table} Summary table of the FSR cluster properties
determined with our isochrone-fitting pipeline (the full table will be published
online only). The table lists the FSR ID number, the cluster type (known open
cluster or new cluster candidate), cluster class (PMS or OC), the distance in
kiloparsec determined using our photometric method in Paper I ($D^{P1}$), our
pipeline ($D^{P2}$) and uncertainty ($ \Delta D^{P2}$); the $H$-band extinction
values calculated from $H-K$ excess using our photometric method in Paper I
($A_H^{P1}$), our pipeline ($A_H^{P2}$) and uncertainty ($\Delta A_H^{P2}$); the
age in $\log(age/yr)$ and uncertainty ($\Delta \log(age/yr)$). Note that $\Delta
A_H^{P2}$ and $\Delta \log(age/yr)$ are only the statistical variations of the
three isochrone fits and do not account for systematical uncertainties due to
the use of solar metallicity isochrones.} \\

\hline
\multicolumn{1}{|c|}{\textbf{FSR}} &
\multicolumn{1}{c|}{\textbf{Type}} &
\multicolumn{1}{c|}{\textbf{Class}} &
\multicolumn{1}{c|}{\textbf{l}} &
\multicolumn{1}{c|}{\textbf{b}} &
\multicolumn{1}{|c|}{\textbf{$D^{P1}$}} &
\multicolumn{1}{c|}{\textbf{$D^{P2}$}} &
\multicolumn{1}{c|}{\textbf{$\Delta D^{P2}$}} &
\multicolumn{1}{|c|}{\textbf{$A_{H}^{P1}$}}  &   
\multicolumn{1}{c|}{\textbf{$A_{H}^{P2}$}}  &
\multicolumn{1}{c|}{\textbf{$\Delta A_{H}^{P2}$}}  &  
\multicolumn{1}{c|}{\textbf{Age}} &
\multicolumn{1}{c|}{\textbf{$\Delta $Age}} \\
\multicolumn{1}{|c|}{\textbf{ID}} &
\multicolumn{1}{|c|}{\textbf{}} &
\multicolumn{1}{|c|}{\textbf{}} &
\multicolumn{1}{c|}{\textbf{[deg]}} &
\multicolumn{1}{c|}{\textbf{[deg]}} &
\multicolumn{1}{c|}{\textbf{[kpc]}} &
\multicolumn{1}{c|}{\textbf{[kpc]}} &
\multicolumn{1}{|c|}{\textbf{[kpc]}} &
\multicolumn{1}{c|}{\textbf{[mag]}}   &
\multicolumn{1}{c|}{\textbf{[mag]}}   &
\multicolumn{1}{|c|}{\textbf{[mag]}}   &
\multicolumn{1}{c|}{\textbf{[$\log(age/yr)$] }} &
\multicolumn{1}{|c|}{\textbf{[$\log(age/yr)$] }} \\ \hline
\endfirsthead

\multicolumn{13}{c}%
{{\bfseries \tablename\ \thetable{} -- continued from previous page}} \\
\hline \multicolumn{1}{|c|}{\textbf{FSR}} &
\multicolumn{1}{c|}{\textbf{Type}} &
\multicolumn{1}{c|}{\textbf{Class}} &
\multicolumn{1}{c|}{\textbf{l}} &
\multicolumn{1}{c|}{\textbf{b}} &
\multicolumn{1}{|c|}{\textbf{$D^{P1}$}} &
\multicolumn{1}{c|}{\textbf{$D^{P2}$}} &
\multicolumn{1}{c|}{\textbf{$\Delta D^{P2}$}} &
\multicolumn{1}{|c|}{\textbf{$A_{H}^{P1}$}}  &   
\multicolumn{1}{c|}{\textbf{$A_{H}^{P2}$}}  &
\multicolumn{1}{c|}{\textbf{$\Delta A_{H}^{P2}$}}  &  
\multicolumn{1}{c|}{\textbf{Age}} &
\multicolumn{1}{c|}{\textbf{$\Delta $Age}} \\
\multicolumn{1}{|c|}{\textbf{ID}} &
\multicolumn{1}{|c|}{\textbf{}} &
\multicolumn{1}{|c|}{\textbf{}} &
\multicolumn{1}{c|}{\textbf{[deg]}} &
\multicolumn{1}{c|}{\textbf{[deg]}} &
\multicolumn{1}{c|}{\textbf{[kpc]}} &
\multicolumn{1}{c|}{\textbf{[kpc]}} &
\multicolumn{1}{|c|}{\textbf{[kpc]}} &
\multicolumn{1}{c|}{\textbf{[mag]}}   &
\multicolumn{1}{c|}{\textbf{[mag]}}   &
\multicolumn{1}{|c|}{\textbf{[mag]}}   &
\multicolumn{1}{c|}{\textbf{[$\log(age/yr)$] }} &
\multicolumn{1}{|c|}{\textbf{[$\log(age/yr)$] }} \\ \hline
\endhead

\hline \multicolumn{13}{|r|}{{Continued on next page}} \\ \hline
\endfoot

\hline \hline
\endlastfoot

0032 &    Known &    OC  &   9.28& -2.53&  2.8 & 1.70 & 0.00 & 0.21 & 0.22 & 0.00 & 9.10 & 0.00 \\
0045 &    Known &    OC  &  12.87& -1.32&  2.2 & 2.60 & 0.00 & 0.20 & 0.32 & 0.00 & 8.50 & 0.00 \\
0071 &    Known &    OC  &  23.89& -2.91&  1.9 & 2.00 & 0.21 & 0.21 & 0.29 & 0.02 & 7.60 & 0.17 \\
0074 &    Known &    OC  &  25.36& -4.31&  3.5 & 5.30 & 0.00 & 0.18 & 0.02 & 0.00 & 9.50 & 0.00 \\
0082 &    Known &    OC  &  27.31& -2.77&  1.1 & 1.60 & 0.07 & 0.09 & 0.14 & 0.03 & 8.60 & 0.09 \\
0089 &    New   &    OC  &  29.49& -0.98&  4.5 & 6.50 & 0.07 & 1.53 & 1.50 & 0.00 & 8.50 & 0.03 \\
0101 &    New   &    OC  &  35.15&  1.75&  3.2 & 1.60 & 0.00 & 1.07 & 1.05 & 0.00 & 9.20 & 0.00 \\
0109 &    Known &    OC  &  37.17&  2.62&  1.7 & 1.50 & 0.03 & 0.52 & 0.59 & 0.01 & 9.00 & 0.00 \\
0111 &    Known &    OC  &  38.66& -1.64&  2.0 & 1.80 & 0.00 & 0.19 & 0.30 & 0.00 & 8.80 & 0.00 \\
0113 &    Known &    OC  &  39.10& -1.68&  1.6 & 2.10 & 0.07 & 0.29 & 0.39 & 0.04 & 8.60 & 0.18 \\
0115 &    Known &    OC  &  40.35& -0.70&  2.4 & 2.20 & 0.00 & 0.80 & 1.10 & 0.00 & 7.10 & 0.00 \\
0122 &    Known &    OC  &  45.70& -0.12&  2.1 & 2.30 & 0.30 & 0.64 & 0.74 & 0.02 & 8.60 & 0.12 \\
0124 &    New   &    OC  &  46.48&  2.65&  3.7 & 1.10 & 0.00 & 0.48 & 0.45 & 0.00 & 9.30 & 0.00 \\
0127 &    Known &    OC  &  48.89& -0.94&  2.6 & 2.90 & 0.18 & 0.60 & 0.64 & 0.01 & 8.20 & 0.02 \\
0133 &    New   &    OC  &  51.12& -1.17&  4.2 & 2.40 & 0.18 & 0.87 & 0.99 & 0.03 & 8.70 & 0.09 \\
0138 &    Known &    OC  &  53.22&  3.34&  2.5 & 3.10 & 0.07 & 0.36 & 0.41 & 0.01 & 9.10 & 0.09 \\
0144 &    Known &    OC  &  56.34& -4.69&  1.9 & 1.70 & 0.20 & 0.02 & 0.05 & 0.03 & 7.80 & 0.10 \\
0154 &    New   &    OC  &  60.00& -1.08&  3.2 & 3.90 & 0.00 & 0.54 & 0.55 & 0.00 & 8.60 & 0.13 \\
0157 &    New   &    OC  &  62.02& -0.70&  2.2 & 1.10 & 0.00 & 0.58 & 0.65 & 0.00 & 6.80 & 0.00 \\
0167 &    New   &    OC  &  65.16& -2.41&  2.4 & 1.60 & 0.15 & 0.42 & 0.43 & 0.03 & 8.70 & 0.12 \\
0168 &    Known &    OC  &  65.53& -3.97&  1.4 & 1.00 & 0.05 & 0.09 & 0.09 & 0.00 & 8.60 & 0.03 \\
0169 &    Known &    PMS &  65.69&  1.18&  2.5 & 2.40 & 0.00 & 0.08 & 0.36 & 0.00 & 7.60 & 0.00 \\
0177 &    Known &    OC  &  67.64&  0.85&  3.1 & 2.80 & 0.00 & 0.15 & 0.30 & 0.00 & 9.20 & 0.00 \\
0186 &    Known &    OC  &  69.97& 10.91&  2.0 & 4.10 & 0.00 & 0.25 & 0.12 & 0.00 & 9.50 & 0.00 \\
0187 &    Known &    OC  &  70.31&  1.76&  4.5 & 5.20 & 0.00 & 0.34 & 0.29 & 0.00 & 8.70 & 0.00 \\
0188 &    New   &    OC  &  70.65&  1.74&  8.3 &10.50 & 1.00 & 0.69 & 0.62 & 0.02 & 8.60 & 0.05 \\
0190 &    New   &    OC  &  70.73&  0.96& 10.2 &11.60 & 0.00 & 1.31 & 1.26 & 0.00 & 8.80 & 0.00 \\
0191 &    New   &    OC  &  70.99&  2.58&  3.8 & 2.40 & 0.37 & 0.56 & 0.59 & 0.04 & 8.50 & 0.18 \\
0195 &    New   &    PMS &  72.07& -0.99&  4.1 & 1.90 & 0.00 & 0.99 & 1.15 & 0.00 & 7.60 & 0.00 \\
0197 &    New   &    OC  &  72.16&  0.30&  3.7 & 1.80 & 0.00 & 0.62 & 0.70 & 0.00 & 8.90 & 0.00 \\
0202 &    Known &    OC  &  73.99&  8.49&  1.5 & 1.80 & 0.10 & 0.02 & 0.07 & 0.00 & 9.20 & 0.05 \\
0205 &    Known &    OC  &  75.24& -0.67&  6.9 & 7.60 & 0.00 & 1.45 & 1.40 & 0.00 & 8.50 & 0.00 \\
0207 &    Known &    PMS &  75.38&  1.30&  2.0 & 1.40 & 0.00 & 0.03 & 0.30 & 0.00 & 7.00 & 0.00 \\
0208 &    Known &    OC  &  75.70&  0.99&  3.2 & 3.40 & 0.20 & 0.42 & 0.56 & 0.02 & 8.20 & 0.12 \\
0214 &    New   &    OC  &  77.71&  4.18&  5.8 & 6.50 & 0.00 & 0.29 & 0.23 & 0.01 & 8.90 & 0.05 \\
0216 &    Known &    OC  &  78.01& -3.36&  1.7 & 1.40 & 0.06 & 0.10 & 0.16 & 0.02 & 8.90 & 0.08 \\
0218 &    Known &    OC  &  78.10&  2.79&  2.7 & 1.00 & 0.00 & 0.37 & 0.38 & 0.00 & 7.40 & 0.00 \\
0231 &    Known &    OC  &  79.57&  6.83&  1.3 & 1.30 & 0.06 &-0.00 & 0.05 & 0.02 & 8.80 & 0.03 \\
0233 &    Known &    OC  &  79.87& -0.93&  3.4 & 1.60 & 0.10 & 1.23 & 1.30 & 0.00 & 9.00 & 0.05 \\
0257 &    New   &    OC  &  83.13&  4.84&  2.8 & 2.30 & 0.00 & 0.26 & 0.24 & 0.00 & 9.50 & 0.00 \\
0267 &    Known &    OC  &  85.68& -1.52&  2.0 & 2.10 & 0.00 & 0.36 & 0.38 & 0.03 & 8.80 & 0.10 \\
0268 &    Known &    OC  &  85.90& -4.14&  3.6 & 3.10 & 0.40 & 0.43 & 0.37 & 0.01 & 9.10 & 0.15 \\
0275 &    New   &    OC  &  87.20&  0.97&  5.1 & 2.40 & 0.00 & 0.52 & 0.40 & 0.00 & 9.30 & 0.00 \\
0276 &    New   &    OC  &  87.32&  5.75&  7.4 & 7.10 & 0.00 & 0.62 & 0.75 & 0.00 & 8.60 & 0.00 \\
0280 &    Known &    OC  &  88.24&  0.26&  4.5 & 4.10 & 0.00 & 0.43 & 0.49 & 0.00 & 9.00 & 0.00 \\
0282 &    New   &    OC  &  88.75&  1.05&  2.6 & 2.70 & 0.00 & 0.45 & 0.56 & 0.02 & 8.80 & 0.09 \\
0285 &    Known &    OC  &  89.62& -0.39&  2.4 & 2.50 & 0.00 & 0.22 & 0.31 & 0.02 & 8.50 & 0.06 \\
0286 &    Known &    OC  &  89.98& -2.73&  1.8 & 1.80 & 0.06 & 0.09 & 0.18 & 0.01 & 8.90 & 0.06 \\
0293 &    New   &    OC  &  91.03& -2.75&  2.3 & 1.40 & 0.00 & 0.06 & 0.15 & 0.00 & 8.30 & 0.00 \\
0294 &    New   &    OC  &  91.27&  2.34&  2.7 & 1.60 & 0.24 & 0.48 & 0.49 & 0.06 & 7.60 & 0.32 \\
0301 &    Known &    PMS &  93.04&  1.80&  4.0 & 2.00 & 0.00 & 0.71 & 1.00 & 0.00 & 7.50 & 0.00 \\
0309 &    Known &    OC  &  94.42&  0.19&  1.7 & 1.60 & 0.09 & 0.18 & 0.30 & 0.02 & 8.20 & 0.06 \\
0320 &    New   &    OC  &  96.38&  1.24&  2.3 & 1.40 & 0.15 & 0.15 & 0.16 & 0.03 & 7.40 & 0.20 \\
0327 &    Known &    OC  &  97.34&  0.45&  3.1 & 1.90 & 0.00 & 0.43 & 0.42 & 0.00 & 7.90 & 0.00 \\
0336 &    New   &    OC  &  99.09&  0.96&  2.5 & 2.30 & 0.12 & 0.37 & 0.52 & 0.02 & 7.30 & 0.38 \\
0342 &    New   &    OC  &  99.76& -2.21&  2.5 & 2.50 & 0.23 & 0.22 & 0.18 & 0.02 & 8.90 & 0.03 \\
0343 &    Known &    OC  &  99.96& -2.69&  2.1 & 2.30 & 0.10 & 0.04 & 0.07 & 0.01 & 8.80 & 0.06 \\
0348 &    Known &    OC  & 101.37& -1.86&  2.0 & 2.10 & 0.00 &-0.01 & 0.01 & 0.01 & 9.00 & 0.03 \\
0349 &    Known &    OC  & 101.41& -0.60&  3.2 & 3.20 & 0.00 & 0.20 & 0.19 & 0.02 & 8.80 & 0.03 \\
0352 &    Known &    OC  & 102.69&  0.80&  2.7 & 1.80 & 0.15 & 0.06 & 0.35 & 0.03 & 7.60 & 0.19 \\
0358 &    New   &    OC  & 103.35&  2.21&  9.9 &10.60 & 0.12 & 1.08 & 1.00 & 0.03 & 8.70 & 0.02 \\
0363 &    Known &    OC  & 104.05&  0.92&  2.9 & 2.90 & 0.00 & 0.23 & 0.18 & 0.00 & 9.10 & 0.00 \\
0373 &    Known &    OC  & 105.35&  9.50&  2.2 & 2.20 & 0.00 & 0.15 & 0.13 & 0.01 & 9.50 & 0.00 \\
0375 &    Known &    OC  & 105.47&  1.20&  2.6 & 2.60 & 0.00 & 0.40 & 0.70 & 0.00 & 7.60 & 0.00 \\
0381 &    New   &    OC  & 106.64& -0.39&  2.3 & 2.20 & 0.00 & 0.16 & 0.16 & 0.03 & 8.80 & 0.06 \\
0382 &    Known &    OC  & 106.64&  0.36&  2.8 & 2.40 & 0.25 & 0.34 & 0.32 & 0.03 & 8.60 & 0.10 \\
0384 &    New   &    OC  & 106.75& -2.95&  2.1 & 1.20 & 0.00 & 0.03 & 0.05 & 0.00 & 7.60 & 0.00 \\
0385 &    New   &    OC  & 106.96&  0.12&  3.0 & 1.90 & 0.00 & 0.40 & 0.35 & 0.00 & 9.00 & 0.00 \\
0388 &    New   &    OC  & 107.32&  5.13&  4.9 & 5.00 & 0.23 & 0.89 & 0.85 & 0.01 & 8.90 & 0.03 \\
0392 &    Known &    OC  & 107.79& -1.02&  2.6 & 2.10 & 0.20 & 0.19 & 0.21 & 0.01 & 8.70 & 0.03 \\
0395 &    Known &    OC  & 108.49& -2.79&  3.0 & 2.50 & 0.21 & 0.20 & 0.26 & 0.01 & 7.70 & 0.13 \\
0396 &    Known &    OC  & 108.51& -0.38&  3.0 & 2.50 & 0.09 & 0.40 & 0.58 & 0.01 & 7.80 & 0.03 \\
0400 &    Known &    OC  & 109.13&  1.12&  4.1 & 2.00 & 0.00 & 0.65 & 1.00 & 0.00 & 7.30 & 0.00 \\
0411 &    Known &    OC  & 110.58&  0.14&  2.9 & 2.10 & 0.00 & 0.31 & 0.26 & 0.00 & 9.00 & 0.00 \\
0412 &    Known &    OC  & 110.70&  0.48&  6.8 & 6.60 & 0.00 & 0.84 & 0.78 & 0.01 & 8.90 & 0.00 \\
0415 &    Known &    OC  & 110.92&  0.07&  2.0 & 1.80 & 0.10 & 0.18 & 0.43 & 0.00 & 7.40 & 0.05 \\
0423 &    New   &    OC  & 111.48&  5.19&  3.2 & 3.10 & 0.12 & 0.42 & 0.43 & 0.03 & 9.20 & 0.03 \\
0430 &    New   &    OC  & 112.71&  3.22&  2.3 & 1.50 & 0.00 & 0.30 & 0.25 & 0.00 & 8.70 & 0.00 \\
0433 &    Known &    OC  & 112.86&  0.17&  2.3 & 1.80 & 0.00 & 0.09 & 0.25 & 0.00 & 8.10 & 0.00 \\
0434 &    Known &    OC  & 112.86& -2.86&  2.2 & 2.10 & 0.00 & 0.23 & 0.25 & 0.00 & 8.40 & 0.03 \\
0444 &    New   &    OC  & 114.51&  2.63&  2.4 & 2.20 & 0.00 & 0.35 & 0.40 & 0.01 & 8.80 & 0.07 \\
0457 &    Known &    OC  & 116.13& -0.14&  1.9 & 1.60 & 0.12 & 0.04 & 0.09 & 0.02 & 8.40 & 0.09 \\
0458 &    Known &    OC  & 116.44& -0.78&  2.2 & 1.80 & 0.06 & 0.07 & 0.14 & 0.01 & 8.00 & 0.08 \\
0461 &    Known &    OC  & 116.60& -1.01&  2.7 & 2.60 & 0.12 & 0.16 & 0.16 & 0.04 & 8.40 & 0.23 \\
0467 &    Known &    OC  & 117.15&  6.49&  3.2 & 3.10 & 0.00 & 0.40 & 0.39 & 0.02 & 9.40 & 0.10 \\
0468 &    Known &    OC  & 117.22&  5.86&  1.8 & 0.80 & 0.00 & 0.23 & 0.32 & 0.00 & 9.00 & 0.00 \\
0475 &    Known &    OC  & 117.99& -1.30&  2.7 & 2.70 & 0.17 & 0.20 & 0.23 & 0.02 & 9.10 & 0.00 \\
0480 &    New   &    OC  & 118.59& -1.09&  6.0 & 5.60 & 0.00 & 0.65 & 0.56 & 0.01 & 8.80 & 0.03 \\
0490 &    Known &    OC  & 119.78&  1.70&  3.6 & 1.50 & 0.00 & 0.38 & 0.40 & 0.01 & 9.10 & 0.00 \\
0491 &    Known &    OC  & 119.80& -1.38&  2.0 & 2.00 & 0.10 & 0.12 & 0.11 & 0.03 & 8.80 & 0.07 \\
0493 &    Known &    OC  & 119.93& -0.09&  2.6 & 2.20 & 0.15 & 0.05 & 0.06 & 0.02 & 8.30 & 0.09 \\
0494 &    New   &    OC  & 120.07&  1.03&  3.2 & 2.90 & 0.00 & 0.21 & 0.25 & 0.02 & 9.40 & 0.07 \\
0496 &    New   &    OC  & 120.26&  1.29&  3.4 & 1.30 & 0.20 & 0.36 & 0.35 & 0.00 & 9.10 & 0.05 \\
0502 &    Known &    OC  & 120.88&  0.51&  2.2 & 2.10 & 0.00 &-0.00 & 0.17 & 0.00 & 8.00 & 0.00 \\
0512 &    Known &    OC  & 122.09&  1.33&  2.6 & 2.20 & 0.12 & 0.11 & 0.21 & 0.03 & 8.80 & 0.06 \\
0519 &    New   &    OC  & 123.05&  1.78&  3.2 & 3.30 & 0.00 & 0.27 & 0.30 & 0.00 & 8.30 & 0.00 \\
0523 &    New   &    OC  & 123.59&  5.60&  2.2 & 2.10 & 0.00 & 0.22 & 0.22 & 0.03 & 9.20 & 0.14 \\
0525 &    Known &    OC  & 124.01&  1.07&  2.3 & 2.00 & 0.06 & 0.14 & 0.23 & 0.02 & 7.90 & 0.10 \\
0528 &    Known &    OC  & 124.69& -0.60&  2.7 & 2.40 & 0.12 & 0.38 & 0.57 & 0.01 & 7.50 & 0.15 \\
0529 &    Known &    OC  & 124.95& -1.21&  2.4 & 1.10 & 0.07 & 0.08 & 0.16 & 0.01 & 8.50 & 0.06 \\
0536 &    New   &    OC  & 126.13&  0.37&  3.0 & 2.20 & 0.27 & 0.45 & 0.52 & 0.04 & 8.50 & 0.13 \\
0540 &    Known &    OC  & 126.64& -4.38&  1.6 & 1.60 & 0.13 &-0.01 & 0.02 & 0.01 & 8.20 & 0.03 \\
0542 &    New   &    OC  & 126.83&  0.38&  4.7 & 4.40 & 0.00 & 0.52 & 0.55 & 0.01 & 9.10 & 0.09 \\
0543 &    Known &    OC  & 127.20&  0.76&  2.7 & 2.40 & 0.00 & 0.24 & 0.24 & 0.03 & 8.90 & 0.12 \\
0548 &    Known &    OC  & 127.75&  2.09&  3.5 & 3.20 & 0.03 & 0.29 & 0.32 & 0.00 & 9.00 & 0.00 \\
0550 &    Known &    OC  & 128.03& -1.80&  1.8 & 1.70 & 0.00 &-0.02 & 0.01 & 0.01 & 8.10 & 0.19 \\
0552 &    Known &    OC  & 128.22& -1.11&  2.4 & 2.00 & 0.00 & 0.08 & 0.17 & 0.00 & 7.80 & 0.00 \\
0554 &    Known &    PMS & 128.56&  1.74&  2.8 & 2.00 & 0.00 & 0.03 & 0.28 & 0.00 & 8.00 & 0.00 \\
0556 &    Known &    OC  & 129.08& -0.35&  1.8 & 1.60 & 0.00 & 0.10 & 0.23 & 0.02 & 8.30 & 0.09 \\
0557 &    Known &    OC  & 129.38& -1.53&  2.5 & 2.00 & 0.15 & 0.11 & 0.18 & 0.01 & 8.40 & 0.08 \\
0559 &    Known &    OC  & 129.51& -0.96&  2.1 & 2.40 & 0.27 & 0.15 & 0.32 & 0.02 & 7.20 & 0.15 \\
0563 &    Known &    OC  & 130.05& -0.16&  4.6 & 5.10 & 0.45 & 0.50 & 0.41 & 0.02 & 8.90 & 0.06 \\
0567 &    Known &    PMS & 130.13&  0.38&  3.2 & 2.20 & 0.00 & 0.24 & 0.35 & 0.00 & 7.70 & 0.00 \\
0574 &    Known &    OC  & 132.42& -6.14&  2.5 & 1.20 & 0.06 & 0.10 & 0.13 & 0.00 & 8.50 & 0.00 \\
0585 &    Known &    OC  & 134.21&  1.07&  4.2 & 3.60 & 0.18 & 0.57 & 0.55 & 0.04 & 8.80 & 0.19 \\
0592 &    Known &    OC  & 135.34& -0.37&  2.8 & 1.10 & 0.00 & 0.27 & 0.22 & 0.00 & 6.70 & 0.00 \\
0594 &    Known &    OC  & 135.44& -0.49&  2.8 & 2.20 & 0.17 & 0.25 & 0.36 & 0.02 & 9.00 & 0.03 \\
0598 &    Known &    PMS & 135.85&  0.27&  1.9 & 2.20 & 0.00 & 0.36 & 0.60 & 0.00 & 7.30 & 0.00 \\
0599 &    Known &    OC  & 136.05& -1.15&  2.3 & 1.90 & 0.12 & 0.21 & 0.24 & 0.06 & 8.70 & 0.26 \\
0603 &    Known &    OC  & 136.31& -2.63&  1.9 & 1.50 & 0.12 & 0.10 & 0.18 & 0.03 & 8.10 & 0.24 \\
0615 &    Known &    PMS & 137.82& -1.75&  2.6 & 1.90 & 0.00 & 0.29 & 0.40 & 0.00 & 7.70 & 0.00 \\
0619 &    Known &    OC  & 138.10& -4.75&  2.5 & 1.40 & 0.00 & 0.11 & 0.10 & 0.00 & 9.20 & 0.00 \\
0623 &    New   &    OC  & 138.62&  8.90&  2.4 & 1.80 & 0.00 & 0.32 & 0.26 & 0.00 & 9.10 & 0.00 \\
0624 &    Known &    OC  & 139.42&  0.18&  5.9 & 5.60 & 0.10 & 0.71 & 0.60 & 0.00 & 9.10 & 0.03 \\
0636 &    Known &    PMS & 143.34& -0.13&  1.8 & 0.80 & 0.00 & 0.25 & 0.35 & 0.00 & 7.70 & 0.00 \\
0639 &    Known &    OC  & 143.78& -4.27&  2.4 & 2.10 & 0.07 & 0.26 & 0.35 & 0.01 & 8.80 & 0.00 \\
0641 &    Known &    OC  & 143.94&  3.60&  2.5 & 1.60 & 0.05 & 0.23 & 0.35 & 0.01 & 8.40 & 0.05 \\
0644 &    Known &    OC  & 145.11& -3.99&  2.5 & 2.00 & 0.09 & 0.24 & 0.21 & 0.02 & 8.30 & 0.03 \\
0645 &    Known &    OC  & 145.92& -2.99&  3.2 & 1.60 & 0.00 & 0.38 & 0.33 & 0.00 & 7.60 & 0.00 \\
0648 &    Known &    OC  & 146.67& -8.92&  1.9 & 2.50 & 0.00 & 0.03 & 0.08 & 0.00 & 8.90 & 0.00 \\
0651 &    Known &    OC  & 147.08& -0.50&  3.9 & 3.50 & 0.00 & 0.68 & 0.75 & 0.00 & 9.20 & 0.00 \\
0652 &    Known &    OC  & 147.52&  5.66&  3.3 & 3.20 & 0.35 & 0.31 & 0.25 & 0.00 & 9.10 & 0.06 \\
0658 &    Known &    OC  & 149.81& -1.01&  3.6 & 3.20 & 0.00 & 0.57 & 0.71 & 0.04 & 8.10 & 0.05 \\
0659 &    Known &    OC  & 149.85&  0.19&  2.7 & 1.40 & 0.00 & 0.23 & 0.28 & 0.03 & 8.60 & 0.10 \\
0677 &    Known &    OC  & 154.84&  2.49&  3.0 & 2.10 & 0.20 & 0.31 & 0.39 & 0.01 & 9.10 & 0.03 \\
0679 &    Known &    OC  & 155.01&-15.32&  1.8 & 0.60 & 0.00 & 0.18 & 0.20 & 0.00 & 8.60 & 0.00 \\
0694 &    Known &    OC  & 158.59& -1.57&  2.7 & 2.60 & 0.13 & 0.23 & 0.29 & 0.03 & 8.80 & 0.07 \\
0705 &    New   &    OC  & 160.71&  4.86&  4.8 & 4.60 & 0.35 & 0.31 & 0.20 & 0.02 & 8.90 & 0.10 \\
0710 &    Known &    OC  & 161.65& -2.01&  4.0 & 3.10 & 0.05 & 0.44 & 0.45 & 0.01 & 9.00 & 0.00 \\
0713 &    Known &    OC  & 162.02& -2.39&  3.1 & 2.70 & 0.00 & 0.27 & 0.30 & 0.00 & 9.10 & 0.00 \\
0718 &    Known &    PMS & 162.27&  1.62&  2.9 & 2.70 & 0.00 & 0.23 & 0.55 & 0.00 & 7.30 & 0.00 \\
0726 &    Known &    OC  & 162.81&  0.66&  5.4 & 5.00 & 0.00 & 0.37 & 0.30 & 0.00 & 8.80 & 0.00 \\
0727 &    New   &    OC  & 162.91&  4.31&  2.9 & 1.70 & 0.19 & 0.16 & 0.21 & 0.02 & 8.80 & 0.09 \\
0728 &    New   &    OC  & 162.92& -6.88&  2.3 & 1.30 & 0.00 & 0.25 & 0.20 & 0.00 & 9.00 & 0.00 \\
0731 &    Known &    OC  & 163.58&  5.05&  5.1 & 4.20 & 0.00 & 0.36 & 0.30 & 0.00 & 9.30 & 0.00 \\
0755 &    Known &    OC  & 168.44&  1.22&  3.3 & 2.80 & 0.18 & 0.13 & 0.20 & 0.01 & 8.50 & 0.03 \\
0769 &    Known &    OC  & 171.90&  0.45&  5.8 & 4.40 & 0.00 & 0.48 & 0.42 & 0.00 & 9.10 & 0.00 \\
0774 &    Known &    OC  & 172.64&  0.33&  2.5 & 1.50 & 0.03 & 0.11 & 0.22 & 0.01 & 8.70 & 0.03 \\
0790 &    New   &    OC  & 173.75& -5.87&  3.4 & 3.20 & 0.00 & 0.26 & 0.18 & 0.00 & 9.20 & 0.00 \\
0792 &    Known &    OC  & 174.10& -8.85&  2.4 & 1.90 & 0.10 & 0.14 & 0.18 & 0.02 & 8.60 & 0.06 \\
0793 &    New   &    OC  & 174.44& -1.86&  4.3 & 4.00 & 0.00 & 0.39 & 0.34 & 0.00 & 8.80 & 0.00 \\
0794 &    Known &    PMS & 174.54&  1.08&  2.0 & 1.20 & 0.00 & 0.06 & 0.20 & 0.00 & 7.30 & 0.00 \\
0802 &    New   &    OC  & 176.17&  6.02&  2.6 & 2.00 & 0.00 & 0.17 & 0.24 & 0.00 & 8.70 & 0.00 \\
0814 &    New   &    OC  & 177.06& -0.41&  3.1 & 1.60 & 0.20 & 0.35 & 0.45 & 0.03 & 8.00 & 0.15 \\
0822 &    Known &    OC  & 179.11&-10.46&  1.8 & 0.80 & 0.03 & 0.10 & 0.18 & 0.02 & 8.60 & 0.12 \\
0825 &    New   &    OC  & 179.32&  1.26&  3.0 & 2.90 & 0.00 & 0.14 & 0.21 & 0.00 & 8.80 & 0.00 \\
0828 &    New   &    OC  & 179.92&  1.75&  5.7 & 5.00 & 0.00 & 0.44 & 0.28 & 0.00 & 8.90 & 0.00 \\
0829 &    Known &    OC  & 179.96& -0.29&  2.8 & 2.10 & 0.00 & 0.27 & 0.21 & 0.00 & 9.20 & 0.00 \\
0847 &    Known &    OC  & 182.74&  0.48&  4.1 & 4.00 & 0.00 & 0.48 & 0.54 & 0.01 & 9.00 & 0.03 \\
0854 &    Known &    OC  & 184.77&-13.51&  1.7 & 1.70 & 0.00 & 0.06 & 0.06 & 0.03 & 9.30 & 0.12 \\
0866 &    New   &    OC  & 186.33& 13.84&  2.1 & 1.30 & 0.00 & 0.01 & 0.05 & 0.00 & 9.10 & 0.00 \\
0867 &    Known &    OC  & 186.37&  1.26&  2.5 & 1.60 & 0.07 & 0.11 & 0.13 & 0.03 & 8.40 & 0.16 \\
0870 &    Known &    PMS & 186.61&  0.15&  2.6 & 1.60 & 0.00 & 0.19 & 0.40 & 0.00 & 7.30 & 0.00 \\
0872 &    Known &    OC  & 186.64&  1.80&  4.4 & 3.10 & 0.00 & 0.25 & 0.19 & 0.00 & 9.40 & 0.00 \\
0881 &    New   &    OC  & 188.06& -2.22&  4.6 & 4.20 & 0.00 & 0.40 & 0.35 & 0.00 & 8.90 & 0.00 \\
0883 &    New   &    OC  & 188.11&  0.15&  2.7 & 2.50 & 0.00 & 0.18 & 0.26 & 0.03 & 8.60 & 0.10 \\
0904 &    New   &    PMS & 191.03& -0.78&  3.1 & 2.00 & 0.00 & 0.34 & 0.43 & 0.00 & 7.30 & 0.00 \\
0942 &    New   &    OC  & 195.58& -3.59&  2.9 & 2.80 & 0.05 & 0.29 & 0.36 & 0.03 & 8.90 & 0.10 \\
0959 &    Known &    OC  & 197.21&  8.92&  2.0 & 4.10 & 0.00 &-0.02 & 0.03 & 0.00 & 8.80 & 0.00 \\
0961 &    Known &    OC  & 197.24& -2.34&  3.0 & 2.90 & 0.00 & 0.13 & 0.17 & 0.01 & 8.80 & 0.03 \\
0971 &    Known &    OC  & 198.04& -5.80&  3.1 & 3.00 & 0.00 & 0.25 & 0.19 & 0.00 & 9.40 & 0.00 \\
0972 &    Known &    OC  & 198.11& 19.65&  1.7 & 1.50 & 0.00 &-0.01 & 0.02 & 0.00 & 9.40 & 0.00 \\
0973 &    Known &    OC  & 199.03&-10.38&  2.3 & 1.70 & 0.00 & 0.17 & 0.15 & 0.00 & 8.50 & 0.00 \\
0982 &    Known &    OC  & 201.79&  2.11&  2.6 & 2.50 & 0.00 & 0.18 & 0.22 & 0.01 & 9.00 & 0.05 \\
0987 &    New   &    OC  & 202.42& -5.12&  3.2 & 2.00 & 0.17 & 0.21 & 0.31 & 0.02 & 8.10 & 0.18 \\
0995 &    Known &    OC  & 203.38& 11.82&  1.8 & 1.80 & 0.07 &-0.02 & 0.03 & 0.01 & 9.00 & 0.07 \\
1002 &    Known &    OC  & 204.37& -1.69&  2.9 & 2.60 & 0.24 & 0.13 & 0.16 & 0.02 & 9.00 & 0.12 \\
1037 &    Known &    OC  & 207.91&  0.30&  2.7 & 1.70 & 0.03 & 0.08 & 0.10 & 0.01 & 8.60 & 0.03 \\
1042 &    Known &    OC  & 208.57& -1.78&  2.5 & 1.20 & 0.10 & 0.24 & 0.26 & 0.02 & 8.20 & 0.12 \\
1055 &    Known &    OC  & 210.57& -2.10&  3.2 & 3.10 & 0.00 & 0.24 & 0.26 & 0.01 & 9.00 & 0.03 \\
1059 &    Known &    OC  & 210.81& -0.24&  2.5 & 1.60 & 0.09 & 0.03 & 0.09 & 0.03 & 8.40 & 0.19 \\
1063 &    New   &    OC  & 211.25& -3.86&  2.9 & 1.80 & 0.00 & 0.15 & 0.18 & 0.00 & 9.10 & 0.00 \\
1070 &    Known &    OC  & 212.16& -3.43&  5.5 & 5.30 & 0.00 & 0.39 & 0.30 & 0.02 & 9.00 & 0.09 \\
1089 &    Known &    OC  & 213.46&  3.30&  2.5 & 2.60 & 0.18 &-0.07 & 0.00 & 0.00 & 9.00 & 0.03 \\
1101 &    Known &    OC  & 214.54& -0.85&  3.1 & 2.00 & 0.00 & 0.07 & 0.06 & 0.00 & 9.30 & 0.00 \\
1104 &    Known &    OC  & 215.31& -2.27&  2.8 & 2.00 & 0.17 & 0.11 & 0.12 & 0.02 & 8.60 & 0.09 \\
1127 &    Known &    OC  & 217.76& -0.69&  2.6 & 1.80 & 0.17 & 0.10 & 0.12 & 0.02 & 8.70 & 0.00 \\
1148 &    Known &    OC  & 219.85& -2.23&  2.7 & 2.40 & 0.13 & 0.08 & 0.20 & 0.01 & 8.20 & 0.12 \\
1165 &    Known &    OC  & 222.04& -5.31&  3.1 & 3.00 & 0.00 & 0.18 & 0.14 & 0.01 & 9.10 & 0.03 \\
1173 &    New   &    OC  & 223.29& -0.48&  3.3 & 2.10 & 0.30 & 0.17 & 0.16 & 0.04 & 8.70 & 0.05 \\
1174 &    Known &    OC  & 223.54& 10.09&  2.9 & 2.80 & 0.00 & 0.15 & 0.10 & 0.00 &10.00 & 0.03 \\
1189 &    Known &    PMS & 224.67&  0.40&  2.0 & 1.20 & 0.00 &-0.01 & 0.10 & 0.00 & 8.00 & 0.00 \\
1206 &    Known &    OC  & 226.59& -2.30&  2.8 & 2.70 & 0.00 & 0.12 & 0.30 & 0.01 & 7.70 & 0.10 \\
1214 &    Known &    OC  & 227.49& -0.56&  5.7 & 4.10 & 0.00 & 0.47 & 0.36 & 0.00 & 9.40 & 0.00 \\
1215 &    Known &    OC  & 227.87&  5.38&  2.0 & 2.00 & 0.00 &-0.06 & 0.00 & 0.00 & 9.10 & 0.03 \\
1222 &    Known &    OC  & 228.95&  4.51&  2.2 & 1.60 & 0.05 &-0.01 & 0.01 & 0.01 & 9.20 & 0.00 \\
1230 &    Known &    OC  & 230.58&  9.95&  1.7 & 2.50 & 0.00 & 0.00 & 0.02 & 0.00 & 9.20 & 0.00 \\
1231 &    Known &    OC  & 230.80&  1.01&  4.9 & 1.70 & 0.00 & 0.39 & 0.28 & 0.00 & 9.50 & 0.00 \\
1240 &    Known &    OC  & 231.80& -0.59&  3.0 & 2.50 & 0.45 & 0.19 & 0.29 & 0.05 & 8.40 & 0.20 \\
1246 &    Known &    OC  & 232.35& -7.30&  2.1 & 2.10 & 0.03 & 0.10 & 0.17 & 0.01 & 8.80 & 0.00 \\
1267 &    New   &    OC  & 235.48&  1.80&  2.4 & 1.90 & 0.20 &-0.01 & 0.05 & 0.01 & 8.90 & 0.00 \\
1271 &    Known &    OC  & 235.99&  5.38&  2.5 & 1.70 & 0.12 & 0.12 & 0.10 & 0.02 & 8.80 & 0.03 \\
1272 &    Known &    OC  & 236.06& -4.62&  3.2 & 1.60 & 0.00 & 0.04 & 0.06 & 0.00 & 9.20 & 0.00 \\
1274 &    Known &    OC  & 236.28&  0.07&  2.2 & 2.20 & 0.00 &-0.00 & 0.09 & 0.02 & 8.30 & 0.12 \\
1284 &    New   &    OC  & 237.94& -5.08&  2.8 & 2.20 & 0.20 & 0.11 & 0.18 & 0.03 & 8.40 & 0.20 \\
1288 &    Known &    OC  & 238.22& -3.34&  2.9 & 1.40 & 0.12 & 0.02 & 0.08 & 0.01 & 8.10 & 0.10 \\
1291 &    Known &    OC  & 238.40& -6.78&  2.1 & 1.00 & 0.00 &-0.01 & 0.05 & 0.00 & 9.00 & 0.00 \\
1299 &    Known &    OC  & 239.93& -4.94&  3.3 & 1.70 & 0.15 & 0.24 & 0.34 & 0.00 & 8.30 & 0.06 \\
1305 &    New   &    OC  & 241.57& -2.51&  2.9 & 2.10 & 0.03 & 0.02 & 0.17 & 0.01 & 8.50 & 0.07 \\
1323 &    Known &    OC  & 245.67& -4.31&  4.2 & 4.20 & 0.00 & 0.18 & 0.13 & 0.00 & 9.00 & 0.00 \\
1325 &    Known &    OC  & 245.91& -1.74&  5.2 & 2.90 & 0.00 & 0.49 & 0.40 & 0.00 & 9.20 & 0.00 \\
1328 &    Known &    OC  & 246.45& -4.46&  2.2 & 2.20 & 0.00 & 0.04 & 0.04 & 0.01 & 8.30 & 0.00 \\
1330 &    Known &    OC  & 246.72& -0.77&  2.2 & 1.60 & 0.06 & 0.12 & 0.20 & 0.01 & 8.60 & 0.03 \\
1333 &    Known &    OC  & 246.79&  3.37&  2.4 & 2.20 & 0.00 & 0.09 & 0.08 & 0.00 & 9.10 & 0.00 \\
1337 &    Known &    OC  & 247.71& -2.52&  2.9 & 1.50 & 0.00 & 0.27 & 0.33 & 0.00 & 9.10 & 0.00 \\
1338 &    Known &    OC  & 247.81&  1.31&  2.4 & 2.50 & 0.07 & 0.08 & 0.09 & 0.04 & 8.80 & 0.07 \\
1340 &    Known &    OC  & 247.95& -4.15&  3.1 & 1.90 & 0.09 & 0.27 & 0.32 & 0.01 & 8.60 & 0.03 \\
1347 &    New   &    OC  & 248.97& -4.12&  3.0 & 1.40 & 0.00 & 0.28 & 0.23 & 0.00 & 9.00 & 0.00 \\
1354 &    Known &    OC  & 249.83&  2.97&  2.2 & 1.60 & 0.03 &-0.02 & 0.05 & 0.01 & 8.60 & 0.03 \\
1358 &    Known &    OC  & 250.44&  1.60&  2.1 & 2.10 & 0.00 &-0.05 & 0.01 & 0.00 & 8.60 & 0.00 \\
1361 &    New   &    OC  & 251.56& -5.00&  2.8 & 1.80 & 0.15 & 0.18 & 0.23 & 0.00 & 9.00 & 0.05 \\
1362 &    Known &    OC  & 251.60&  6.65&  1.9 & 1.70 & 0.00 & 0.01 & 0.05 & 0.00 & 9.00 & 0.00 \\
1373 &    Known &    OC  & 254.57&  6.08&  1.3 & 2.80 & 0.09 &-0.01 & 0.02 & 0.01 & 8.90 & 0.03 \\
1375 &    Known &    OC  & 255.61&  3.98&  2.3 & 2.30 & 0.00 & 0.15 & 0.09 & 0.00 & 9.10 & 0.00 \\
1384 &    Known &    OC  & 257.27&  4.27&  2.0 & 2.10 & 0.13 & 0.15 & 0.16 & 0.01 & 9.20 & 0.03 \\
1386 &    Known &    OC  & 257.99& -1.00&  5.5 & 5.70 & 0.00 & 0.78 & 0.69 & 0.03 & 8.90 & 0.07 \\
1387 &    New   &    OC  & 258.12& -1.33&  4.6 & 4.70 & 0.00 & 0.69 & 0.63 & 0.00 & 8.80 & 0.00 \\
1388 &    Known &    OC  & 258.50&  2.30&  3.0 & 3.20 & 0.13 & 0.29 & 0.23 & 0.01 & 9.20 & 0.06 \\
1392 &    Known &    OC  & 258.87& -3.33&  6.3 & 2.90 & 0.00 & 0.72 & 0.57 & 0.00 & 9.20 & 0.00 \\
1393 &    Known &    OC  & 259.06&  2.00&  3.5 & 3.80 & 0.00 & 0.38 & 0.27 & 0.00 & 8.90 & 0.00 \\
1399 &    New   &    OC  & 259.95&  2.06&  2.1 & 2.60 & 0.00 & 0.33 & 0.33 & 0.00 & 9.30 & 0.00 \\
1404 &    Known &    OC  & 261.53&  3.76&  2.6 & 2.70 & 0.00 & 0.26 & 0.22 & 0.01 & 9.20 & 0.00 \\
1415 &    New   &    OC  & 263.74& -1.81&  9.1 & 9.30 & 0.00 & 0.84 & 0.78 & 0.01 & 9.10 & 0.07 \\
1420 &    Known &    OC  & 264.09& -5.51&  2.4 & 3.10 & 0.00 & 0.14 & 0.15 & 0.00 & 9.10 & 0.00 \\
1424 &    New   &    PMS & 264.19&  0.18&  2.8 & 1.10 & 0.00 & 0.48 & 0.47 & 0.00 & 7.30 & 0.00 \\
1430 &    New   &    OC  & 264.65&  0.08&  7.0 & 7.10 & 0.00 & 1.32 & 1.30 & 0.01 & 8.50 & 0.00 \\
1433 &    Known &    PMS & 264.81& -2.91&  3.0 & 1.30 & 0.00 & 0.10 & 0.20 & 0.00 & 7.50 & 0.00 \\
1436 &    New   &    PMS & 264.91& -2.87&  3.4 & 2.00 & 0.00 & 0.18 & 0.30 & 0.00 & 7.00 & 0.00 \\
1444 &    Known &    OC  & 265.80& -5.01&  3.4 & 2.40 & 0.00 & 0.17 & 0.12 & 0.00 & 9.00 & 0.00 \\
1450 &    New   &    OC  & 266.94& -0.37&  5.8 & 5.90 & 0.00 & 1.00 & 0.95 & 0.00 & 8.80 & 0.00 \\
1452 &    New   &    OC  & 267.60& -2.09&  3.0 & 2.10 & 0.20 & 0.21 & 0.27 & 0.03 & 8.50 & 0.15 \\
1458 &    Known &    OC  & 268.65&  3.21&  2.1 & 1.50 & 0.00 & 0.16 & 0.19 & 0.00 & 9.20 & 0.00 \\
1460 &    New   &    OC  & 269.13& -0.19&  3.7 & 3.50 & 0.15 & 0.83 & 0.94 & 0.01 & 9.00 & 0.00 \\
1472 &    Known &    OC  & 270.76&  3.22&  2.4 & 2.50 & 0.00 & 0.37 & 0.37 & 0.00 & 8.90 & 0.00 \\
1480 &    Known &    OC  & 272.50&  2.87&  2.1 & 2.00 & 0.00 & 0.13 & 0.15 & 0.00 & 9.30 & 0.00 \\
1482 &    Known &    OC  & 273.13& -0.77&  2.3 & 2.20 & 0.07 & 0.20 & 0.27 & 0.01 & 8.10 & 0.07 \\
1487 &    Known &    OC  & 273.82&-15.89&  1.2 & 0.40 & 0.00 & 0.01 & 0.06 & 0.01 & 7.90 & 0.06 \\
1502 &    Known &    OC  & 277.11& -0.81&  2.3 & 1.50 & 0.00 & 0.09 & 0.12 & 0.00 & 9.00 & 0.00 \\
1508 &    New   &    OC  & 278.51& -0.61&  2.9 & 2.70 & 0.09 & 0.27 & 0.41 & 0.02 & 7.90 & 0.09 \\
1515 &    Known &    OC  & 279.48&  0.15&  2.6 & 2.80 & 0.07 & 0.12 & 0.14 & 0.01 & 9.20 & 0.00 \\
1520 &    New   &    OC  & 280.21&  0.07&  2.4 & 1.70 & 0.00 & 0.16 & 0.13 & 0.00 & 9.40 & 0.00 \\
1521 &    New   &    OC  & 280.44& -1.62&  5.8 & 5.90 & 0.00 & 0.53 & 0.47 & 0.00 & 9.20 & 0.00 \\
1522 &    New   &    OC  & 280.71&  0.12&  2.4 & 1.80 & 0.00 & 0.26 & 0.25 & 0.00 & 8.80 & 0.00 \\
1526 &    Known &    OC  & 282.06& -2.40&  2.5 & 2.10 & 0.32 & 0.09 & 0.04 & 0.00 & 9.00 & 0.03 \\
1530 &    New   &    OC  & 282.34& -1.07&  6.5 & 6.60 & 0.00 & 1.00 & 1.30 & 0.00 & 6.80 & 0.00 \\
1533 &    Known &    OC  & 283.01&  0.44&  2.1 & 2.10 & 0.10 & 0.18 & 0.16 & 0.03 & 8.50 & 0.15 \\
1534 &    Known &    OC  & 283.14& -1.46&  2.7 & 2.30 & 0.26 & 0.10 & 0.25 & 0.02 & 8.30 & 0.04 \\
1537 &    Known &    OC  & 283.85& -3.69&  2.7 & 2.70 & 0.00 & 0.01 & 0.02 & 0.01 & 9.00 & 0.00 \\
1540 &    Known &    OC  & 284.62&  0.04&  1.9 & 1.70 & 0.15 &-0.00 & 0.10 & 0.01 & 8.20 & 0.08 \\
1544 &    Known &    OC  & 285.34& -8.82&  1.4 & 1.30 & 0.06 &-0.01 & 0.03 & 0.02 & 8.70 & 0.06 \\
1545 &    Known &    OC  & 285.87&  0.08&  1.4 & 1.50 & 0.07 &-0.12 & 0.01 & 0.01 & 7.50 & 0.07 \\
1551 &    Known &    PMS & 287.40& -0.34&  1.8 & 1.30 & 0.00 & 0.02 & 0.30 & 0.00 & 7.30 & 0.00 \\
1558 &    Known &    OC  & 288.69&  0.43&  2.2 & 2.10 & 0.06 & 0.20 & 0.19 & 0.01 & 8.00 & 0.07 \\
1559 &    Known &    OC  & 289.16&  0.31&  5.6 & 3.30 & 0.00 & 0.54 & 0.40 & 0.00 & 9.40 & 0.00 \\
1562 &    Known &    OC  & 289.52& -0.40&  2.3 & 2.20 & 0.10 & 0.20 & 0.25 & 0.00 & 8.60 & 0.00 \\
1564 &    Known &    OC  & 289.90& -5.57&  2.1 & 1.70 & 0.03 & 0.03 & 0.12 & 0.03 & 8.60 & 0.13 \\
1565 &    Known &    OC  & 290.19&  2.88&  1.9 & 2.00 & 0.00 &-0.02 & 0.00 & 0.00 & 9.70 & 0.03 \\
1575 &    Known &    OC  & 291.21& -0.16&  2.0 & 1.80 & 0.19 & 0.04 & 0.09 & 0.02 & 7.90 & 0.09 \\
1576 &    Known &    PMS & 291.64& -0.51&  3.5 & 1.00 & 0.00 & 0.66 & 0.90 & 0.00 & 7.70 & 0.00 \\
1582 &    New   &    OC  & 292.38& -1.82&  2.0 & 1.80 & 0.15 & 0.15 & 0.26 & 0.03 & 7.90 & 0.20 \\
1586 &    New   &    OC  & 292.84& -1.20&  4.4 & 4.10 & 0.25 & 0.61 & 0.64 & 0.02 & 8.90 & 0.10 \\
1587 &    Known &    OC  & 292.92& -2.41&  1.5 & 2.00 & 0.25 & 0.08 & 0.19 & 0.03 & 8.30 & 0.10 \\
1588 &    Known &    OC  & 293.21&  0.58&  3.8 & 4.00 & 0.20 & 0.25 & 0.17 & 0.01 & 8.90 & 0.07 \\
1589 &    Known &    OC  & 294.11& -0.03&  1.1 & 1.60 & 0.17 &-0.10 & 0.04 & 0.03 & 7.70 & 0.20 \\
1590 &    Known &    OC  & 294.38&  6.18&  1.8 & 1.70 & 0.03 &-0.05 & 0.04 & 0.01 & 9.20 & 0.02 \\
1591 &    New   &    OC  & 294.52& -1.09&  5.6 & 5.80 & 0.10 & 0.86 & 0.85 & 0.00 & 8.70 & 0.00 \\
1596 &    Known &    OC  & 295.79& -0.21&  3.2 & 2.20 & 0.00 & 0.46 & 0.47 & 0.00 & 8.90 & 0.00 \\
1600 &    Known &    OC  & 297.52& -1.76&  3.8 & 3.30 & 0.07 & 0.48 & 0.40 & 0.01 & 9.00 & 0.07 \\
1603 &    New   &    OC  & 298.22& -0.51&  2.2 & 2.40 & 0.25 & 0.25 & 0.21 & 0.01 & 9.20 & 0.06 \\
1611 &    Known &    OC  & 299.32&  4.56&  1.8 & 1.90 & 0.07 & 0.11 & 0.13 & 0.02 & 9.30 & 0.07 \\
1614 &    Known &    OC  & 299.76&  0.86&  1.9 & 1.60 & 0.30 &-0.05 & 0.14 & 0.01 & 8.30 & 0.10 \\
1615 &    Known &    OC  & 300.11& -0.67&  3.5 & 3.50 & 0.00 & 0.58 & 0.60 & 0.02 & 9.20 & 0.03 \\
1624 &    Known &    OC  & 301.50&  2.20&  2.9 & 3.40 & 0.00 & 0.27 & 0.18 & 0.00 & 9.00 & 0.00 \\
1627 &    Known &    OC  & 301.71& -5.53&  3.5 & 1.50 & 0.00 & 0.32 & 0.26 & 0.00 & 9.70 & 0.00 \\
1633 &    Known &    OC  & 303.22&  2.47&  1.4 & 1.40 & 0.00 &-0.00 & 0.10 & 0.01 & 7.50 & 0.03 \\
1637 &    Known &    OC  & 303.63& -2.08&  2.4 & 2.30 & 0.03 & 0.31 & 0.30 & 0.02 & 8.90 & 0.07 \\
1644 &    New   &    OC  & 305.51& -4.32&  2.2 & 1.70 & 0.05 & 0.19 & 0.19 & 0.03 & 8.50 & 0.25 \\
1655 &    Known &    OC  & 307.74&  1.56&  2.0 & 1.60 & 0.15 & 0.12 & 0.14 & 0.01 & 8.40 & 0.07 \\
1670 &    Known &    OC  & 310.84&  0.16&  5.1 & 2.50 & 0.00 & 1.11 & 1.19 & 0.00 & 8.50 & 0.00 \\
1679 &    Known &    OC  & 314.72& -0.30&  4.2 & 3.40 & 0.40 & 0.53 & 0.60 & 0.05 & 8.80 & 0.05 \\
1686 &    New   &    OC  & 316.00& -0.29&  5.0 & 1.70 & 0.00 & 0.97 & 0.97 & 0.00 & 9.30 & 0.00 \\
1704 &    Known &    OC  & 325.80& -2.97&  2.7 & 2.00 & 0.18 & 0.22 & 0.21 & 0.01 & 9.10 & 0.06 \\
1706 &    Known &    OC  & 326.01& -1.93&  1.5 & 1.30 & 0.00 & 0.08 & 0.07 & 0.02 & 8.90 & 0.07 \\
1716 &    New   &    OC  & 329.79& -1.59&  6.4 & 5.40 & 0.00 & 0.89 & 0.79 & 0.03 & 9.10 & 0.07 \\
1723 &    New   &    OC  & 333.03&  5.85&  1.1 & 1.10 & 0.03 &-0.02 & 0.00 & 0.00 & 8.80 & 0.03 \\
1726 &    Known &    OC  & 334.55&  1.09&  3.0 & 2.30 & 0.00 & 0.37 & 0.38 & 0.00 & 9.10 & 0.03 \\
1730 &    Known &    OC  & 335.47& -6.24&  1.4 & 0.90 & 0.00 & 0.05 & 0.08 & 0.00 &10.00 & 0.00 \\
1738 &    Known &    OC  & 340.11& -7.88&  1.4 & 1.00 & 0.06 & 0.12 & 0.15 & 0.00 & 9.00 & 0.03 \\

\end{longtable}

\end{center}

\end{appendix}

\label{lastpage}
\end{document}